%% file: lsmgraph.tex
\documentclass[acmsmall]{acmart} % camera ready

\pdfoutput=1 % to submit to arxiv

\AtBeginDocument{%
  }

\setcopyright{acmlicensed}
\acmJournal{PACMMOD}
\acmYear{2024} \acmVolume{2} \acmNumber{6 (SIGMOD)} \acmArticle{243} \acmMonth{12}\acmDOI{10.1145/3698818}
\usepackage{booktabs}
\usepackage{subcaption}                                           
\usepackage{endnotes,microtype,xspace,graphicx,fancyvrb,multirow} 
\usepackage[linesnumbered, ruled, vlined,resetcount]{algorithm2e} 
\usepackage{algpseudocode}
\usepackage[normalem]{ulem}
\newcommand{\red}[1]{\textcolor[rgb]{1,0,0.4}{#1}}

\newcommand{\eat}[1]{}

\usepackage{balance}
\input{commands}

\usepackage{comment}

\newcommand{\sys}{\texttt{LSMGraph}\xspace}
\newcommand{\mgs}{MemGraph\xspace}
\newcommand{\dst}{\textit{dst}\xspace}

\newcommand{\offset}{\textit{offset}\xspace}

\newcommand{\marker}{\textit{marker}\xspace}

\newcommand{\mmap}{\textit{mmap}\xspace}

\usepackage{newfloat}
\usepackage{multirow}
\usepackage{bbding} % 邮箱符号\Envelope
\begin{document}

%%
%% The "title" command has an optional parameter,
%% allowing the author to define a "short title" to be used in page headers.
\title{LSMGraph: A High-Performance Dynamic Graph Storage System with Multi-Level CSR}

%%
%% The "author" command and its associated commands are used to define
%% the authors and their affiliations.
%% Of note is the shared affiliation of the first two authors, and the
%% "authornote" and "authornotemark" commands
%% used to denote shared contribution to the research.
\eat{
\author{Ben Trovato}
\authornote{Both authors contributed equally to this research.}
\email{trovato@corporation.com}
\orcid{1234-5678-9012}
\author{G.K.M. Tobin}
\authornotemark[1]
\email{webmaster@marysville-ohio.com}
\affiliation{%
  \institution{Institute for Clarity in Documentation}
  \city{Dublin}
  \state{Ohio}
  \country{USA}
}

\author{Lars Th{\o}rv{\"a}ld}
\affiliation{%
  \institution{The Th{\o}rv{\"a}ld Group}
  \city{Hekla}
  \country{Iceland}}
\email{larst@affiliation.org}

\author{Aparna Patel}
\affiliation{%
 \institution{Rajiv Gandhi University}
 \city{Doimukh}
 \state{Arunachal Pradesh}
 \country{India}}

\author{Huifen Chan}
\affiliation{%
  \institution{Tsinghua University}
  \city{Haidian Qu}
  \state{Beijing Shi}
  \country{China}}

\author{Charles Palmer}
\affiliation{%
  \institution{Palmer Research Laboratories}
  \city{San Antonio}
  \state{Texas}
  \country{USA}}
\email{cpalmer@prl.com}

\author{John Smith}
\affiliation{%
  \institution{The Th{\o}rv{\"a}ld Group}
  \city{Hekla}
  \country{Iceland}}
\email{jsmith@affiliation.org}

\author{Julius P. Kumquat}
\affiliation{%
  \institution{The Kumquat Consortium}
  \city{New York}
  \country{USA}}
\email{jpkumquat@consortium.net}
} % end eat

\author{Song Yu}
\email{yusong@stumail.neu.edu.cn}
\author{Shufeng Gong}
\email{gongsf@mail.neu.edu.cn}
\affiliation{%
  \institution{ Northeastern University}
  % \city{Shenyang Shi}
  \country{China}
}

% \author{Shufeng Gong}
% \affiliation{%
%   \institution{ Northeastern University}
%   % \city{Shenyang Shi}
%   \country{China}
% }
% \email{gongsf@mail.neu.edu.cn}

\author{Qian Tao}
\email{qian.tao@alibaba-inc.com}
\author{Sijie Shen}
\email{shensijie.ssj@alibaba-inc.com}
\affiliation{%
  \institution{Alibaba Group}
  % \city{Beijing Shi}
  \country{China}
}

% \author{Sijie Shen}
% \affiliation{%
%   \institution{Alibaba Group}
%   % \city{Beijing Shi}
%   \country{China}
% }
% \email{shensijie.ssj@alibaba-inc.com}

\author{Yanfeng Zhang}
\authornote{Yanfeng Zhang is the corresponding author.}
\affiliation{%
  \institution{ Northeastern University}
  % \city{Shenyang Shi}
  \country{China}
}
\email{zhangyf@mail.neu.edu.cn}

\author{Wenyuan Yu}
\affiliation{%
  \institution{Alibaba Group}
  % \city{Beijing Shi}
  \country{China}
}
\email{wenyuan.ywy@alibaba-inc.com}

\author{Pengxi Liu}
\email{liupengxi@stumail.neu.edu.cn}
\author{Zhixin Zhang}
\email{yinshi@stumail.neu.edu.cn}
\author{Hongfu Li}
\email{lihongfu@stumail.neu.edu.cn}
\affiliation{%
  \institution{ Northeastern University}
  % \city{Shenyang Shi}
  \country{China}
}

% \author{Zhixin Zhang}
% \affiliation{%
%   \institution{ Northeastern University}
%   % \city{Shenyang Shi}
%   \country{China}
% }
% \email{yinshi@stumail.neu.edu.cn}

% \author{Hongfu Li}
% \affiliation{%
%   \institution{ Northeastern University}
%   % \city{Shenyang Shi}
%   \country{China}
% }
% \email{lihongfu@stumail.neu.edu.cn}

\author{Xiaojian Luo}
\affiliation{%
  \institution{Alibaba Group}
  % \city{Beijing Shi}
  \country{China}
}
\email{lxj193371@alibaba-inc.com}

\author{Ge Yu}
\affiliation{%
  \institution{ Northeastern University}
  % \city{Shenyang Shi}
  \country{China}
}
\email{yuge@mail.neu.edu.cn}

\author{Jingren Zhou}
\affiliation{%
  \institution{Alibaba Group}
  % \city{Beijing Shi}
  \country{China}
}
\email{jingren.zhou@alibaba-inc.com}

%%
%% By default, the full list of authors will be used in the page
%% headers. Often, this list is too long, and will overlap
%% other information printed in the page headers. This command allows
%% the author to define a more concise list
%% of authors' names for this purpose.
\renewcommand{\shortauthors}{Song Yu et al.}

%%
%% The abstract is a short summary of the work to be presented in the
%% article.
% \begin{abstract} 
% xxxx
% \end{abstract}
%%%%%%%%%%%%%%%%%%%%%%%%%%%%%%%%%%%%%%%%%%%%%%%%%%%%%%%%%%%%
% abstract 
\input{0-abs}


%%
%% Keywords. The author(s) should pick words that accurately describe
%% the work being presented. Separate the keywords with commas.
% \keywords{Do, Not, Us, This, Code, Put, the, Correct, Terms, for,

% \received{20 February 2007}
% \received[revised]{12 March 2009}
% \received[accepted]{5 June 2009}

% Articles V2mod223-V2mod254 use
\received{April 2024}
\received[revised]{July 2024}
\received[accepted]{August 2024}

%%
%% This command processes the author and affiliation and title
%% information and builds the first part of the formatted document.
\maketitle

%%%%%%%%%%%%%%%%%% all section %%%%%%%%%%%%%%%%
% \section{Introduction}
% ACM's consolidated article template, introduced in 2017, provides a

% \section{Template Overview}
% As noted in the Introduction,

\input{1-intro}

\input{2-bg_opt}

\input{3-overview}

\input{4-design}

\input{6-eval}

\input{7-related}

\input{8-concl}

%%%%%%%%%%%%%%%%%%%%%%%%%%%%%%%%%%%%%%%%%%%%%%%%%%%%%%%%

%%
%% The acknowledgments section is defined using the "acks" environment
%% (and NOT an unnumbered section). This ensures the proper
%% identification of the section in the article metadata, and the
%% consistent spelling of the heading.
\begin{acks}
We thank the anonymous reviewers for their constructive comments and suggestions.
The work is supported by the National Key R\&D Program of China (2023YFB4503601), 
the National Natural Science Foundation of China (U2241212, 62072082, and 62202088), 
the 111 Project (B16009), 
the Distinguished Youth Foundation of Liaoning Province (2024021148-JH3/501), 
the Joint Funds of Natural Science Foundation of Liaoning Province (2023-MSBA-078), 
and the Fundamental Research Funds for the Central Universities (N2416011).
\end{acks}

%%
%% The next two lines define the bibliography style to be used, and
%% the bibliography file.
\bibliographystyle{ACM-Reference-Format}
\bibliography{lsmgraph}

\end{document}

%% file: commands.tex
\usepackage{latexsym}
\usepackage{amsfonts}
\usepackage{amsmath}
\usepackage{amssymb}
\usepackage{color}
\usepackage{epsfig}
\usepackage{xspace}
\usepackage{graphicx}
\usepackage{cleveref}
\usepackage{balance}
\usepackage{hhline}
\usepackage{float}
\usepackage{xcolor}

\usepackage{epsfig}
\usepackage{multirow}
\usepackage{url}

\usepackage{enumitem}
\setlist{topsep=0pt,noitemsep} \setitemize[1]{label=$\circ$}
\newcommand{\ei}{\end{itemize}\vspace{1ex}}

\newcommand{\ee}{\end{enumerate}\vspace{1ex}}
\newcommand{\beqn}{\begin{eqnarray}}
\newcommand{\eeqn}{\end{eqnarray}}

\newcommand{\stitle}[1]{\vspace{1.2ex}\noindent{\bf #1}}
\newcommand{\etitle}[1]{\vspace{1.2ex}\noindent{\em\underline{#1}}}

\newcommand{\ie}{\emph{i.e.,}\xspace}
\newcommand{\eg}{\emph{e.g.,}\xspace}

\renewcommand{\smallskip}{\vspace{0.6ex}}

\newcounter{ccc}

\newcommand{\eop}{\hspace*{\fill}\mbox{$\Box$}}     % End of proof
\newcounter{example}%[section]
\renewcommand{\theexample}{\arabic{example}}
\newenvironment{example}{
        \vspace{1.2ex}
        \refstepcounter{example}
        {\noindent\bf Example \theexample:}}{
        \eop\vspace{1.2ex}}

\newlist{myitemize}{itemize}{3}
\setlist[myitemize,1]{label=$\circ$,leftmargin=3.8ex}
\setlist[myitemize,2]{label=$\bullet$,leftmargin=3.8ex}
\setlist[myitemize,3]{label=$\diamond$, leftmargin=3.2ex}

\newcommand{\mei}{\end{myitemize}\vspace{0.6ex}}

\usepackage{tikz}

%% file: 0-abs.tex
\begin{abstract} 
The growing volume of graph data may exhaust the main memory. It is crucial to design a disk-based graph storage system to ingest updates and analyze graphs efficiently.
However, existing dynamic graph storage systems suffer from read or write amplification and face the challenge of optimizing both read and write performance simultaneously.
To address this challenge, we propose LSMGraph, a novel dynamic graph storage system that combines the write-friendly LSM-tree and the read-friendly CSR. 
It leverages the multi-level structure of LSM-trees to optimize write performance while utilizing the compact CSR structures embedded in the LSM-trees to boost read performance.
% To ingest graph updates fast, we design a memory structure MemGraph to cache graph updates efficiently. 
LSMGraph uses a new memory structure, MemGraph, to efficiently cache graph updates and uses a multi-level index to speed up reads within the multi-level structure.
Furthermore, LSMGraph incorporates a vertex-grained version control mechanism to mitigate the impact of LSM-tree compaction on read performance and ensure the correctness of concurrent read and write operations. 
Our evaluation shows that LSMGraph significantly outperforms state-of-the-art (graph) storage systems on both graph update and graph analytical workloads.
\end{abstract}

%% file: 1-intro.tex
\section{Introduction}
\label{sec:intro}

Real-time analysis of large-scale dynamic graph data has become a key requirement in various fields, extensively applied in recommendation systems~\cite{recomPPR,recomPPRVLDB,recommendationWangHZZZL18,recommenderTanLZYZH20}, fraud detection~\cite{fraudDetection,fraudDetectionMLP,graphscope}, community management~\cite{LPIncCommunity,lpCommunity}, network monitoring~\cite{graphone,networkAnalysis,networkdetection} and so on.
Designing an efficient dynamic graph storage system capable of rapid data storage and effective real-time graph analysis is increasingly crucial and challenging. %as it requires the system to 
Such a system must not only support fast real-time graph analysis, but also efficiently ingest updates, especially in scenarios where graphs are updated frequently, such as social networks, e-commerce, and smart city management~\cite{graphone, graphanalysis, SubMillisecond,livegraph}.

\begin{figure}[tbph]
  \centering
  \includegraphics[width=3.0in]{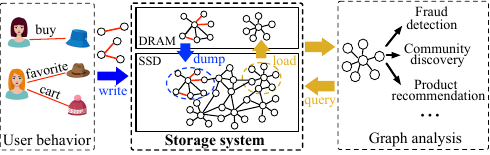}
  \vspace{-0.05in}
  \caption{An example of a graph storage system working in an e-commerce platform, where users or items represent vertices and the interactions between users and items are regarded as edges of the graph.
  } 
  \label{fig:realScenario}
  % \vspace{-0.25in}
\end{figure}

\stitle{Real-world Example}. 
Consider a graph storage system that stores the user-item graphs on a large-scale e-commerce platform, such as Alibaba Taobao~\cite{taobao}. 
Fig. \ref{fig:realScenario} shows a toy example of the graph storage system,
where users or items represent vertices, and the interactions between users and items, such as favorite and buy, are regarded as edges of the graph. 
The system should maintain graph updates and support graph analysis tasks such as product recommendation and fraud detection.
However, the massive number of users and their frequent behaviors bring huge challenges to the graph storage system.
First, Taobao has approximately 400 million daily active users~\cite{taobaoDAU}, each generating an average of 10 behavioral data records per day~\cite{taobaodata}.
This means that Taobao generates approximately 46,000 behavioral data records per second.
\textit{Therefore, the storage system must support fast data ingestion. 
}
Second, the average size of each behavioral data is approximately 31 bytes~\cite{taobaodata}. 
Given the high data generation rate, a 1 TB RAM will be exhausted in less than 9 days, \ie $1 \text{ TB} / (4 \times 10^8 \text{ users} \times 10 \text{ bh./day} \times 31 \text{ bytes})\approx 8.9 \text{ days}$, where bh. refers to customer behaviors such as buying and favoriting.
\textit{Therefore, the storage system needs to efficiently digest updates using limited memory and persist them on disks.} 
In this paper, we default to using SSDs as persistent storage, as they provide a good balance between storage capacity, price, and performance. 
Third, graph data needs to be analyzed using graph algorithms, such as using PageRank for product recommendations~\cite{recomPPR, recomPPRVLDB} and Label Propagation for fraud detection~\cite{fraudDetection,fraudDetectionMLP}.
\textit{Therefore, the storage system needs to support fast queries for these graph analysis algorithms.}

\stitle{Existing Disk-based Dynamic Graph Storages}.
A naive method
is to extend the existing memory-based dynamic graph storage systems~\cite{livegraph, risGraph} to disk-based ones by the utilization of automatic memory and disk mapping tools like 
\mmap \cite{mmap}.
When updates occur on the graph data stored on disk, these systems automatically load the graph data into memory using \mmap to perform in-place graph updates. 
However, this strategy leads to significant read/write amplification, although it makes data operations easier.
This is because even if only one edge is updated, \mmap must read the entire data page from the disk into memory and then write back the entire page after the updates.

On the other hand, there are some dynamic graph storage systems specifically designed for disks, 
\eg LLAMA~\cite{llama} and GraphSSD~\cite{graphssd}, which use Compressed Sparse Rows (CSR)~\cite{csr}
to improve read performance since CSR is friendly to both random and sequential access to
graphs due to its efficient offset index and compactness.
However, the compactness of CSR leads to a large amount of data movement when inserting new data, which decreases graph update performance.
Although LLAMA overcomes data movement by generating new CSRs in batches, a large number of CSR fragments are generated as updates continuously arrive.
To access all neighbors of a vertex, it is required to read data from multiple CSRs, which results in a lot of random I/Os and decreases graph query performance. 

Finally, some graph databases, such as NebulaGraph \cite{NebulaGraph} and Dgraph \cite{dgraph2023}, leverage Log-Structured Merge-tree (LSM-tree) \cite{lsm-tree} designed primarily for key-value stores to improve write performance. 
NebulaGraph \cite{NebulaGraph} takes each vertex or edge as a key, and the properties of the vertex or edge as the value. 
However, when Nebula accesses all the edges of a vertex, the non-contiguous storage of these edges results in a large number of random reads, which decreases read performance.
Dgraph \cite{dgraph2023} takes each vertex as a key, and the associated edges as the value. 
However, when Dgraph updates an edge of a vertex, it requires updating the whole edge block (as a value), which results in severe write amplification.
This deficiency is more pronounced for vertices with more edges.  
In summary, these works suffer from problems in reading or writing as they ignore the properties of the graphs.

\stitle{Dilemma}. 
It is challenging to design a graph storage system for persistent storage since there is a dilemma in optimizing read
and write performance.
In terms of optimizing write performance, 
a log structure is preferred with only sequential writes (\eg LSM-tree \cite{lsm-tree}). However, due to the random arrival order of updated edges, the edges of a vertex can be dispersed throughout the whole log-structured storage, which decreases the read performance when accessing all the edges of a vertex. 
In terms of optimizing read performance, a compact and sorted structure (\eg CSR \cite{csr}) is preferred with efficient indexing and sequential reads. 
However, it requires significant overhead due to the data movement to maintain compactness when edges are inserted or deleted.

\stitle{Insight}. 
LSM-tree \cite{lsmtree} is a %compact and sorted 
log structure known for its excellent write performance,
implemented by caching recent writes in memory and then performing sequential writes to persistent storage.
On the contrary, CSR \cite{csr} is a widely used graph storage structure with excellent graph query performance due to its continuous storage of edges for each vertex and efficient vertex/edge index.
It is worthwhile to design a novel graph storage structure by combining LSM-tree and CSR so that the structure can simultaneously leverage high performance in the writing of LSM-tree and high performance in the reading of CSR. 

\stitle{Our solution}. Based on the above insight, we propose \sys, a novel dynamic graph storage system that takes advantage of both LSM-tree and CSR.
From a memory perspective, 
we design an efficient memory cache structure for graph data, \textit{MemGraph}, which can not only rapidly ingest graph updates but also effectively persist LSM-style data structures to disk.
From a disk perspective,
we propose organizing the graph data into multiple levels similar to the LSM-tree, and each level maintains a part of the graph data in CSR format.
When the storage of the $i$-th level is full, a \textit{compaction} is performed in the background, asynchronously merging the CSR of the $i$-th level into the next level.
In this way, the online continuous data movement overhead of CSR is replaced by the offline periodical compaction overhead of LSM-tree. 
It is noteworthy that although the edges of a vertex are stored continuously in each CSR, they may span multiple CSRs across different levels, 
which results in the need to locate their position in multiple files and decreases read performance.
To enhance read performance, we design a \textit{multi-level index} that records the positions of each vertex's edges on multiple levels to avoid many random searches.
Additionally, we design a \textit{vertex-grained version control mechanism} to mitigate the compaction overhead and allow concurrent read/write operation during compaction. 

To sum up, our contributions\eat{ of this paper} are summarized as follows.

\begin{itemize}[leftmargin=*]
    \item[$\bullet$] A novel dynamic graph storage system \sys that leverages the write performance of LSM-tree and the read performance of CSR (Section \ref{sec:overview}).
    
   \item[$\bullet$] 
   A memory cache structure MemGraph that manages the cached graph updates efficiently before flushing them to disk (Section \ref{sec:design:memgraph}).
   
    \item[$\bullet$] 
    A multi-level index that supports fast reading of a vertex's edges from multiple CSRs on different levels (Section \ref{sec:design:multi-level}).

    \item[$\bullet$]  A vertex-grained version control mechanism that maximizes read and write performance during CSR compaction, while simultaneously guaranteeing the correctness of reading and writing (Section \ref{sec:design:version}). 

    \item[$\bullet$] 
    A high-performance implementation and a comprehensive evaluation to verify the efficiency of our \sys (Section \ref{sec:expr}). 
    Experiments show that for graph update, 
    \sys achieves an average speedup of 
    36.12$\times$ over LiveGraph, 
    2.85$\times$ over LLAMA, 
    and 8.07$\times$ over RocksDB. 
    For graph analysis,
    \sys achieves an average speedup of 
    24.4$\times$ over LiveGraph, 
    3.1$\times$ over LLAMA, 
    30.8$\times$ over RocksDB, 
    and 6.6$\times$ over MBFGraph.
\end{itemize}

%% file: 2-bg_opt.tex
\section{Background}
\label{sec:bg}

In this section, we first introduce graph workloads and their requirements for graph storage and then review the design of CSR and LSM-trees.

\subsection{Graph and Graph Workloads}
\label{sec:bg:graphWorkload}

\stitle{Graph}.
Given a graph $G = (V, E)$, where $V$ is a finite set of vertices and $E \subseteq V \times V$ is a set of edges.
The property of each edge $(u,v)\in E$ is denoted by $p_{u,v}$, which indicates properties such as weight and label, and can be empty.

\stitle{Graph Update}. 
Graph update is the operation of modifying or updating graph data. It includes adding and deleting vertices/edges and modifying properties of vertices/edges.
The graph storage system should support the rapid storage of updates and adjust the structure of the graph according to the updates.

\stitle{Graph Analysis}.
Graph analysis algorithms, such as PageRank, SSSP, Label Propagation, etc., require discovering relationships between entities in graph data and some valuable insights.
The most basic operation in the graph analysis algorithm is to scan all edges of a vertex \cite{sortledton,livegraph}. 
Therefore, to achieve efficient graph analysis\eat{ algorithms}, the graph storage structure should support fast neighbor scanning.
Besides, algorithms such as triangle counting for graph pattern matching require scanning of ordering neighbors for fast intersections~\cite{emptyHeaded,edgeFrame,subgraphQueries}. 
Therefore, the graph storage system should support both ordered and unordered scanning neighbors.

\subsection{CSR and LSM-tree}
\label{sec:bg:csrLSM}

We next introduce two storage structures, CSR and LSM-tree, and analyze their read and write performance in terms of storing graphs.

\begin{figure*} %[tbp]
    \centering
    \begin{minipage}{0.56\textwidth}
        \begin{center}
            \includegraphics[width=2.9in]{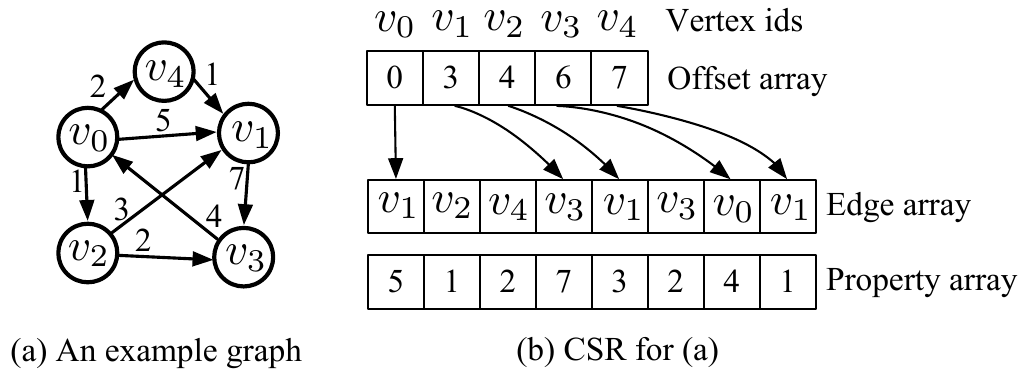}
            % \vspace{0.05in}
            \caption{An example graph and its CSR.}
            \label{fig:csr}
        \end{center}
    \end{minipage}%
    \ 
    \hspace{0.01in}
    \begin{minipage}{0.41\textwidth}
        \begin{center}
            \includegraphics[width=2in]{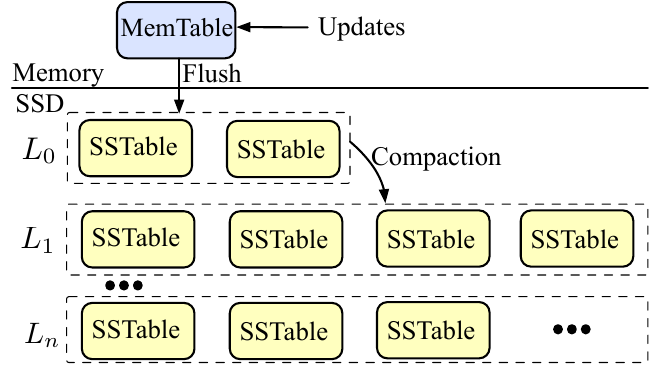}
            \vspace{-0.05in}
            \caption{A classic implementation of LSM-tree.}
            \label{fig:lsm-tree}
        \end{center}
    \end{minipage}
    \hspace{0in}
  \vspace{-0.15in}
\end{figure*}

\stitle{CSR}.
Compressed Sparse Row (CSR)~\cite{csr} is a type of compressed sparse matrix, which stores the non-zero elements of the matrix in multiple dense arrays.
As shown in Fig. \ref{fig:csr}, CSR utilizes three arrays to store a graph: offset array, edge array, and property array. 
\begin{itemize}[leftmargin=*]
\vspace{1.3ex}
    \item \noindent The offset array stores the offset of the first edge of each vertex in the edge array.
    \noindent\item The edge array stores the associated edges of each vertex, \ie destination vertices, in a continuous manner.
    \noindent\item The property array stores the properties corresponding to each edge in the edge array, such as weights.
    \vspace{1.3ex}
\end{itemize}
It stores sparse graphs in a compact and contiguous format, providing excellent spatial locality and space efficiency, making it widely adopted in many graph systems \cite{grape,gemini,galois,ligra,graphchi,graphwalker}.

\etitle{Read Performance Analysis}.
In CSR, edges are stored in the edge array and indexed by an offset array. %Therefore, 
When reading a vertex's edges, the first edge's offset in the edge array can be obtained from the offset array using the vertex ID, requiring only one memory I/O, \ie $O(1)$ memory I/O, as the offset array is usually cached in memory in the existing graph systems. 
Then, the edge data is loaded from the disk with one disk I/O according to the offset, \ie $O(1)$ disk I/O.
Additionally, to obtain edge properties, an extra disk I/O is required. 
The primary reason for storing edges and properties separately in CSR is to accommodate different types of graph analysis algorithms, as some algorithms only need the graph's topology without edge properties, \eg BFS.
%Table \ref{tab:ReadWriteAnalysis} summarizes these costs.

% \begin{figure}[tbp]
%     \centering
%     % \vspace{-0.4in}
%     \includegraphics[width=2.4in]{figures/csr.pdf}
%     \vspace{-0.15in}
%     \caption{An example graph and its CSR representation.}
%     \label{fig:csr}
%      \vspace{-0.25in}
% \end{figure}

% \begin{figure}[tbp]
%     \centering
%     % \vspace{-0.4in}
%     \includegraphics[width=1.8in]{figures/lsm-tree.pdf}
%     \vspace{-0.1in}
%     \caption{A classic implementation of LSM-tree.}
%     \label{fig:lsm-tree}
%     \vspace{-0.2in}
% \end{figure}

\etitle{Write Performance Analysis}.
Due to the compactness of CSR,
the cost of inserting and deleting edges in CSR 
is expensive because these operations may result in a significant data movement to maintain compactness.
In the worst-case workload, inserting an edge into a CSR requires moving all edges/properties in the edge/property array and modifying all offsets in the offset array. 
To mitigate the overhead, we partition the edge array by the block size $B$, ensuring that each I/O operation can move a maximum of $B$ edges.
Here, $B$ donates the size of a block, \ie the number of edges in one block.
The amortized disk I/O times for data movement is 
% $O(\frac{|E|}{2B})$, 
$O(\frac{|E|}{B})$, 
and the 
% amortized 
memory I/O times for modifying the offset is 
% $O(\frac{|V|}{2})$.
$O(|V|)$.

\stitle{LSM-tree}.
The Log-Structured Merge-tree (LSM-tree)~\cite{lsmtree} is 
designed to improve the performance of writing ordered data to disk. 
Fig. \ref{fig:lsm-tree} shows a classic implementation of LSM-tree, which has been widely adopted in numerous key-value stores \cite{leveldbcode,rocksdb,MyRocks,bigtable,xEngine,cassandra}.
It consists of an in-memory \textit{MemTable} and disk-based \textit{Sorted String Tables (SSTable)} organized on multiple levels.
The Memtable is typically implemented with a \textit{skip list} that supports fast insertions and queries.
The SSTable is a \textit{fixed-size immutable} structure stored on disk in an ordered manner, where the data is sorted based on the keys.

\eat{
\begin{table}[tbp]
\centering
\caption{Term Definitions \red{need?}}
\small
\begin{tabular}{|c|l|}
\hline
\textbf{Term} & \textbf{Definition} \\ \hline
$E$ & set of all edges \\ \hline
$V$ & set of all vertices \\ \hline
$P$ & buffer size (edges) \\ \hline
$T$ & size ratio between adjacent levels\\ \hline
$L$ & number of LSM-tree levels \\ \hline
$B$ & number of edges that fit into a storage block \\ \hline
$d$ & degree of a vertex \\ \hline
\end{tabular}
\label{tab:terms}
\end{table}
}

In an LSM-tree, data is typically organized using key-value pairs. When storing graph data, 
it is better to take the edge $e(src, dst)$ as the key and the property of the edge as the value \cite{livegraph}, where the src and dst represent the IDs of the edge's source and destination vertex. 

% \begin{figure}[tbp]
%     \centering
%     % \vspace{-0.4in}
%     \includegraphics[width=1.8in]{figures/lsm-tree.pdf}
%     \vspace{-0.1in}
%     \caption{
%     %Workflow of LSM-tree. 
%     A classic implementation of LSM-tree.
%     }
%     \label{fig:lsm-tree}
%     \vspace{-0.2in}
% \end{figure}

\etitle{Write Performance Analysis}.
%%%%%   Monkey/Dostoevsky/LSM-base/Chucky
%%%%%   Chucky: https://dl.acm.org/doi/pdf/10.1145/3448016.3457273
%%%%%   Monkey: https://dl.acm.org/doi/pdf/10.1145/3035918.3064054
%%%%%   Dostoevsky: https://dl.acm.org/doi/pdf/10.1145/3183713.3196927
%%%%%   LSM-base: https://link.springer.com/article/10.1007/s00778-019-00555-y
LSM-tree first caches graph updates in Memtable, \ie a skip list. 
Inserting an edge incurs $O(\log (P))$ %$\approx$ log($P$) 
memory I/Os, since  
a binary search is performed, where $P$ is the maximum number of edges that can be cached in MemTable.
When the MemTable is full, it is flushed to the disk in SSTable format.
LSM-tree employs a multi-level structure to store data on disk. 
The SSTable flushed from memory is stored at $L_0$. 
When the number of SSTables at $L_i$ reaches a threshold, the LSM-tree compacts these SSTables from $L_i$ to $L_{i+1}$.
The capacity of each level grows exponentially by a factor of $T$, \ie the capacity of $L_i$ is $P\cdot T^{i}$ edges.
To facilitate query performance, we only consider the leveling compaction strategy \cite{Dostoevsky,Monkey}, in which
compaction can be regarded as a multi-way merge that sorts data by key so that the edges of each vertex are stored contiguously at the current level.
Notably, the compaction process always occurs asynchronously in the background, thus hiding the overhead of data movement from writing tasks.
The I/O cost of updating an edge is amortized through the subsequent merge operations in which the updated edge participates.
In a worst-case workload, all updates reach the largest level $L= \left\lceil \log_T \left( \frac{|E|}{P} \cdot \frac{T - 1}{T} \right) \right\rceil$.
This means that each edge will be merged across all levels without being discarded by a more recent edge at a smaller level \cite{Dostoevsky}.
Before reaching capacity, the edges at each level undergo an average of $O(T)$ merge operations, and across $O(L)$ levels, there are a total of $O(T \cdot L)$ merge operations.
We divide the total number of merge operations by the block size $B$ since every disk I/O during a merge operation copies $B$ edges from the original SSTables to the new SSTable. Thus, the amortized disk I/O times for one update is $O(\frac{L \cdot T}{B})$ I/O.

\begin{table}[tbp]
\centering
\caption{Comparison of Memory and Disk I/O Complexities for Different Structures. 
\eat{
$E$ represents the set of all edges, 
$V$ represents the set of all vertices, 
$P$ denotes the cache size in edges, 
$T$ is the size ratio between adjacent levels, 
$L$ is the number of LSM-tree levels, 
$B$ refers to the number of edges that fit into a storage block, and
$d$ is the degree of a vertex.
}}
% \vspace{-0.15in}
\label{tab:ReadWriteAnalysis}
\small
% \footnotesize
\setlength{\tabcolsep}{2.0pt} % Adjust column spacing as needed
\begin{tabular}{l|c|c|c|c}
\hline
\multirow{2}{*}{\textbf{Structure}} & \multicolumn{2}{c|}{\textbf{Read}} & \multicolumn{2}{c}{\textbf{Write}} \\ \cline{2-5} 
 & \textbf{Memory I/O} & \textbf{Disk I/O} & \textbf{Memory I/O} & \textbf{Disk I/O} \\ \hline
CSR & $O(1)$ & $O(1)$ & $O(|V|)$ & $O(\frac{|E|}{B}))$ \\ \hline
LSM-tree & $O(\log(|E|))$ & $O(L)$ & $O(\log(P))$ & $O\left(\frac{L \cdot T}{B}\right)$ \\ \hline
LSMGraph & $O(1)$ & $O(L)$ & $O(\log(d))$ & $O\left(\frac{L \cdot T}{B}\right)$ \\ \hline
\end{tabular}
\label{tab:io_comparison}
\vspace{-0.15in}
\end{table}

From the above analysis, we can see that LSM-tree improves writing performance in three aspects. 1) It caches and reorders random arrival edges and writes them to the disk in an append-only manner. 2) The expensive online data movement is replaced by the offline compaction periodically. 
3) It employs a multi-level structure to organize disk data to mitigate write amplification of compaction.

\etitle{Read Performance Analysis}.
In LSM-tree, the edges of a vertex may be distributed across both the MemTable in memory and the SSTables on disk.
For edges in MemTables, query operations require $O(\log (P))$ times memory I/O.
For edges in SSTables, given that each level's SSTables and their internal data are ordered, enable binary search queries to locate the relevant data block that contains the queried vertex in $O(\log (|E|))$ memory I/Os.
Then a Bloom filter \cite{bloomfilter,Chucky,Monkey} is used to detect whether the found data block needs to be loaded from the disk.
For example, RocksDB's blocked Bloom filters require only one memory I/O per query \cite{Chucky}.
Since $P$ is usually much smaller than $|E|$, the total memory I/O complexity is $O(\log(|E|))$.
If the Bloom filter query returns true, one disk I/O is needed to load the data block. 
In the worst-case workload, each level contains edges for the vertex, which requires loading multiple data blocks. Therefore, the total disk I/O is $O(L)$.

Compared to graph-aware structures like CSR, LSM-trees have poorer read performance in two aspects. 
1) Reading a vertex's edges involves a large amount of memory I/O, and in the worst case, requires disk I/Os equal to the maximum number of levels.
2) LSM-tree's SSTables store data as key-value pairs and organize them in fixed quantities without considering the semantics of graphs.
In contrast, CSR indexes edges by source vertex ID and stores all edges compactly by source vertex.
It was reported that LSM-tree is on average 55 times slower than CSR in reading edges of vertices \cite{livegraph}.

Some graph databases \cite{NebulaGraph,dgraph2023,bytegraph} use LSM-tree based storage systems to store graph data, such as NebulaGraph \cite{NebulaGraph}, Dgraph \cite{dgraph2023} and ByteGraph \cite{bytegraph}. 
In NebulaGraph \cite{NebulaGraph}, keys store vertices or edges, while values store the properties of vertices or edges. 
Dgraph \cite{dgraph2023} stores vertices as keys and edges of a vertex as the value. 
Additionally, ByteGraph \cite{bytegraph} stores the edges of each vertex as a tree structure in memory, with each node of the tree stored as a key-value pair. 
It should be noted that these systems use existing LSM-tree based systems as their underlying storage, such as RocksDB \cite{rocksdb} for NebulaGraph and ByteGraph, and Badger \cite{Badger} for Dgraph. This approach, which treats LSM-tree based system as a black box, does not incorporate the structural properties and workload
characteristics of graphs, and thus still suffers from the same read performance issues analyzed above for LSM-trees.

%% file: 3-overview.tex
\section{Opportunities and Overview}
\label{sec:overview}

\subsection{Combination Opportunities}
\label{sec:opportunity}

As we analyzed in Section \ref{sec:bg:csrLSM},
CSR has been widely adopted in static graph systems \cite{grape,gemini,galois,ligra,graphchi,graphwalker}, as it optimizes read performance through compact storage and efficient indexing. 
In contrast, LSM-tree, with its multi-level structure that replaces random write with periodical sequential write, has become the preferred storage structure for databases that prioritize write performance \cite{leveldbcode,rocksdb,MyRocks,bigtable,xEngine,cassandra}. 

Given the advantages and disadvantages of CSR and LSM-tree, neither structure is ideally suited for supporting both high read and write performance on its own. 
Fortunately, we now identify two opportunities to combine their advantages for this work.

\stitle{Opportunity \#1: Utilize LSM-tree to hide the movement overhead of CSR}. 
As analyzed in Section \ref{sec:bg:csrLSM}, 
when inserting or deleting edges, CSR may move enormous data to keep compactness, which is unacceptable as it results in significant I/O overhead on disk. The design of the LSM-tree that turns random writes into sequential writes and online data movement into offline periodic compactions, which is the reason why the LSM-tree has good write performance, can be used to manage CSR to avoid the overhead of online data movement.

\stitle{Opportunity \#2: Utilize CSR to accelerate graph reading}. 
LSM-tree is non-graph-aware. For some frequent query operations, such as reading all edges of a vertex or finding k-hop neighbors of a vertex, will result in a large number of random reads. Suppose the graph data at each level in the LSM-tree is stored in CSR format. In that case, the efficient index and graph-aware storage of CSR, in which the edges of each vertex are stored continuously and ordered on the disk, will be exploited when reading the graph data to improve read performance.

Although there are opportunities to combine their advantages to build new structures that provide high read and write performance, it is non-trivial to efficiently embed CSR into the LSM-tree, as several challenges need to be addressed.

\stitle{Write}.
Like LSM-tree, to provide high write performance, it first caches updates in memory, then flushes them to disk. 
To efficiently flush the cached updates to disk,
the edges of each vertex should be stored continuously. 
A simple method is to reserve a fixed contiguous space for each vertex in memory to store its edges. 
However, the number of edges of each vertex varies, typically following a power-law distribution.
This means that when a high-degree vertex's space is full, it is necessary to expand the capacity and move data, while vertices with few edges will waste memory resources. 
Therefore, designing a cache structure that can be efficiently updated and flushed to disk presents a challenge.

\stitle{Read}. 
While the multi-level structure of LSM-tree provides efficient write performance, it complicates the query operations. Since the edges of a vertex may be distributed across multiple levels, extensive lookups are required at each level to locate the data in existing LSM-tree designs, resulting in poor read performance. Designing an index to accelerate the location of a vertex's edges in a multi-level structure is challenging.

\stitle{Write-Read Concurrency}. 
In an LSM-tree architecture, the cached data in memory is continuously flushed, disk files are dynamically compacted and new files are generated. 
If a query operation encounters deleted or duplicate data, it will lead to incorrect results. 
It is a challenge to design an efficient version management strategy that ensures each vertex reads the correct version of the data while maintaining high parallelism.

Based on the above two opportunities, we design an efficient disk-based dynamic graph storage system \sys and overcome the mentioned challenges.

\subsection{Overview}

\begin{figure}[tbp]
% \vspace{-0.4in}
  \centering
  \includegraphics[width=3in]{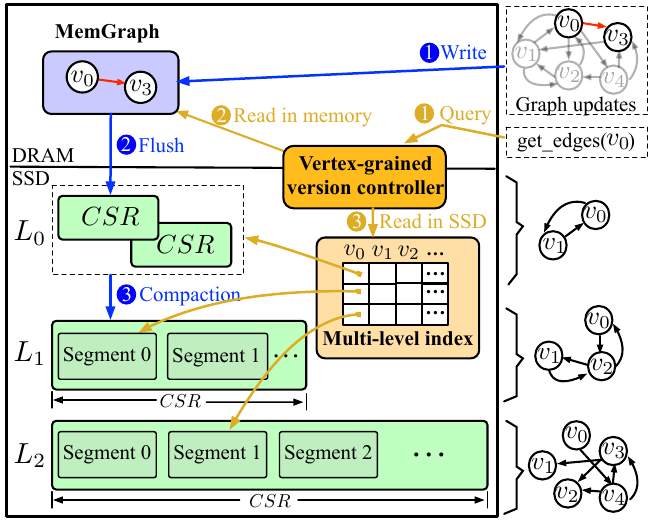}
  % \vspace{0.05in}
  \vspace{-0.1in}
  \caption{Overall architecture of \sys.
   }
  \Description[]{}
  \label{fig:overview}
  \vspace{-0.15in}
\end{figure}

\eat{The system is designed to meet the application scenarios described in Section \ref{sec:intro}, and employs a series of innovative designs to enhance its performance.}
The overall architecture of \sys is shown in Fig. \ref{fig:overview}.

\stitle{Architecture}.
As shown in Fig. \ref{fig:overview}, the system consists of four components, \ie \mgs, multi-level CSR, multi-level index, and vertex-grained version controller.
MemGraph is an efficient in-memory structure to cache recently arrived graph updates efficiently. 
CSRs organized on different levels are stored on disk and the multi-level index is built to read the edges of vertices from CSRs of different levels efficiently. 
Vertex-grained version controller is used to support concurrent read/write operations and mitigate the impact of compaction on read/write performance.

\stitle{Workflow}. 
The read and write workflow of \sys is as follows.

\etitle{Write}. In \sys, when the graph update arrives, \eg the added edges or deleted edges, they are firstly inserted into the \mgs (Section \ref{sec:design:memgraph}). 
As graph updates continue to be inserted, once the \mgs reaches its capacity limit, it is flushed to $L_0$ on disk with the CSR format. 
Note that, to improve write performance, the \mgs is directly written to $L_0$ without compacting with CSRs at $L_0$.
When the 
number of CSR files at $L_0$ reaches its limit, compaction is triggered to merge multiple CSR files at $L_0$ and $L_1$ into a new CSR, which is then written \eat{on}to $L_1$ (Section \ref{sec:disk:csr}). 
There is only one CSR on each level, 
except for $L_0$, and it can be divided into several segments. 
Similarly to LSM-tree, when $L_i$ reaches its capacity limit, compaction is triggered to merge CSR segments and write them to $L_{i+1}$.
The multi-level index is also updated after compaction (Section \ref{sec:disk:index}).

\etitle{Read}.
When performing graph analysis algorithms, there are different query requests, such as reading a vertex/edge, reading all edges of a vertex, etc. 
We take reading all edges of vertex $v_0$ as an example to demonstrate the query process, as it covers the most common query flow.
When the query operation is triggered to read the edges of vertex $v_0$, \sys utilizes vertex-grained version controller to obtain a set of accessible data, denoted as \textit{version}, which includes \mgs and the index of $v_0$ in the multi-level index.
\sys first searches and reads data from the \mgs recorded in the version. 
Then, according to the index in the version, \sys can obtain the position information of $v_1$'s edges on each level, and read the corresponding data based on this information (Section \ref{sec:design:version}).

%% file: 4-design.tex
\section{System Design%{\sys}
}
\label{sec:design}

\begin{table}[tbp] % "t!" means "force to the top"
  \footnotesize
  \caption{Degree distribution of cached data in LSM-tree memory cache structure.}
  \label{tab:degreeDistribution}
  % \vspace{0.05in}
  \centering
  \small
  \setlength{\tabcolsep}{4.5pt} % Adjust column spacing as needed
  \begin{tabular}{lccccccc}
  \toprule
    Degree & [1,2] & [3,4] & [5,8] & [9,16] & $>$16 \\
    \midrule
    IT-2004	\cite{it-2004} & 96.35\% & 2.85\% & 0.64\% & 0.15\% & 0.01\% \\
    UK-2007 \cite{uk-2007} & 93.53\% & 3.55\% & 1.59\% & 0.74\% & 0.58\% \\
    Twitter \cite{twitter} & 89.45\% & 5.66\% & 2.94\% & 1.19\% & 0.76\% \\
    Friendster \cite{friend} & 94.57\% & 4.58\% & 0.81\% & 0.04\% & 0.00\% \\
  \bottomrule
  \end{tabular}
  \vspace{-0.15in}
\end{table}

In this section, we present the details of each core component in \sys.

\subsection{Memory Cache Structure}
\label{sec:design:memgraph}

In LSM-tree, the MemTable is used to cache random writes and enhance write performance by converting these into sequential disk writes, making it a crucial component of the system.
However, MemTable is primarily designed for key-value data and does not consider the characteristics of graph data and workloads, rendering it inadequate for graph storage requirements. Therefore, designing a memory cache structure that supports fast graph updates and queries is critical for improving both the read and write performance of \sys. 

We present the following two important observations before describing the detailed design of %\mgs.
\sys's memory cache structure.

\etitle{Observation 1}.
For LSM-tree based works such as RocksDB\cite{rocksdb}, the MemTable is typically implemented as a skip list. 
A popular implementation \cite{livegraph,NebulaGraph} for storing graphs using these works takes an edge as a key-value pair, where the vertex ID pair (\ie <src, dst>) of the edge serves as the unique key, and the edge property as the value. 
The edges are usually sorted by source vertex ID (\ie src) in ascending order and destination vertex ID (\ie dst) in second ascending order.
For any inserted edge, the MemTable needs to search for the correct position in the skip list based on the key, and then insert it to maintain the overall key-sorted order of the skip list. 
Since the MemTable stores all edges of all vertices in the same skip list, inserting and querying edges for a vertex require a complexity of $O(\log n)$, where $n$ is the total number of stored edges.
Furthermore, as introduced in Section \ref{sec:bg:graphWorkload}, a common operation in graph analysis tasks is scanning all the edges of a vertex. 
However, scanning all edges of a vertex in the MemTable is not efficient due to the discontinuous memory storage of the skip list.
Since MemTable is not optimized for the traits of graph data and workloads, simply applying MemTable's design would lead to poor insertion and query performance.

\etitle{Observation 2}.
In most scenarios, \eg shopping networks, social networks, and web networks, graphs often follow a power-law distribution \cite{yao2023ragraph,powerlaw1barabasi1999emergence,powerlaw2faloutsos1999power}. 
Table \ref{tab:degreeDistribution} illustrates the degree distribution of vertices in the memory cache structure of LSM-tree (with a size of 64MB default in RocksDB) when storing the real-world graph datasets (see Table \ref{tab:dataset} for details).
It can be observed that approximately 95\% of the vertices have no more than 4 edges, while less than 1\% of the vertices have more than 16 edges. 
This inspires us to consider the characteristics of data distribution when designing memory cache structures.

\begin{figure}[tbp]
  \centering
  % \vspace{-0.4in}
  \includegraphics[width=3.0in]{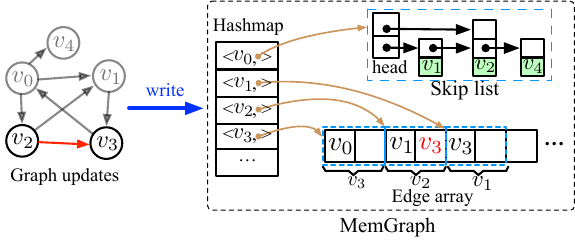}
  % \vspace{0.05in}
  \vspace{-0.15in}
  \caption{An example of \mgs.
  }
  \label{fig:MGS}
  \vspace{-0.15in}
\end{figure}

\stitle{\mgs}.
Based on the above observations, we design a new memory cache structure for graph data,  
Fig. \ref{fig:MGS} illustrates the structure of {\mgs}. 
The \mgs utilizes a hashmap to store vertex IDs and the address of their first edge. 
On one hand, the hashmap provides approximate array-like query performance. 
On the other hand, since \mgs stores only a part of the graph data at a time, the vertices are sparse, and using a hashmap as an index can save more memory than using an array.
The \mgs employs an edge array and a skip list to store the neighbors of low-degree and high-degree vertices respectively. This design choice is based on the consideration of the power-law nature of graphs, as we mentioned in Observation 2. 
Storing low-degree vertices in an array since the array provides good spatial locality, which enables fast scanning and quicker flushing to disk. 
However, storing high-degree vertices in the array may cause a large amount of data movement when inserting edges. 
Thus, for high-degree vertices, we store their edges in a skip list, which has a better insertion performance and logarithmic query complexity.

Specifically, for vertices with low degrees, we store their edges in the edge array. Different from the edge array in CSR, the edge array in \mgs is divided into multiple equally sized segments, and each segment only stores the edges of one vertex. 
Note that, to prevent data movement, the assignment of each segment on the edge array does not follow the order of the source vertices. Instead, segments are assigned based on the order in which the edges arrive, \eg the first segment is allocated to the source vertex of the first arriving edge.
In Fig. \ref{fig:MGS}, with the segment size set to 2, the unique outgoing edge of vertex $v_3$ is stored in the first segment of the edge array, as $e(v_3, v_0)$ is the first arriving edge.
For vertices with edges larger than the segment size, \mgs inserts their edges into a skip list.
As shown in Fig. \ref{fig:MGS}, vertex $v_0$ has three edges that exceed the size of the array segment, so its edges are stored in a skip list.

\subsection{Multi-level CSR}
\label{sec:design:multi-level}

In this section, we first introduce the structure of multi-level CSR, which incorporates vertex-aware compaction. We then introduce the multi-level index to facilitate efficient reading of graph data.

\subsubsection{Compact Multi-level CSR}
\label{sec:disk:csr}

As we mentioned in Section \ref{sec:bg:csrLSM}, the multi-level structure of the LSM-tree effectively mitigates the write amplification of compaction, thus, we also employ a multi-level structure to organize CSRs on disk.
As shown in Fig. \ref{fig:overview}, unlike the LSM-tree, where there are multiple fixed-size SSTables, 
we ensure that there is only one CSR at $L_i$ ($i > 0$), while multiple CSRs are allowed at $L_0$ to flush \mgs more quickly.
For each level, the capacity, similar to an LSM-tree, grows exponentially by a factor of $T$, with $T=10$ by default.
We divide the CSR at $L_i$ ($i > 0$) into multiple segments, each stored in a separate file.
When determining the size of each CSR segment, we try to balance the size of each segment as much as possible while ensuring that the edges of each vertex are assigned to the same segment.
This approach ensures load balancing during compaction and avoids accessing multiple segments when reading the edges of a vertex within a level.
However, if a vertex has many edges, the segment containing these edges may become very large, leading to more severe read and write amplification during compaction. To address this, 
we allow each vertex with many edges to be stored in a separate segment.
By segmentation, when the CSR's size at $L_i$ ($i > 0$) exceeds the capacity of $L_i$, only one of the segments (denoted as $S_j$) is compacted to $L_{i+1}$ without affecting the other data of the CSR at $L_i$. 
Meanwhile, only a few segments at $L_{i+1}$ are involved in compaction, \ie segments at $L_{i+1}$ whose edges have the same source vertex with edges in $S_j$. 
Regarding $L_0$, its multiple CSRs originate from the flushing of \mgs, and their vertex ranges almost always overlap. If each CSR is compacted to the next level individually, it would cause many identical files at the next level to be compacted multiple times. Therefore, we require all overlapping CSRs of $L_0$ to be compacted to the next level in a single compaction to reduce write amplification.

Compared to the simple graph data stored in Fig. \ref{fig:csr}, the CSR file requires more information to be persisted on disk, such as metadata to support data recovery and lookup, as well as timestamps for snapshot isolation. 
Additionally, in each CSR segment, there is only a part of vertices' edges. Employing the original CSR storage format would result in space wastage, which requires allocating arrays with the size of the number of vertices. 
These differences prompt us to design a new CSR storage format for CSR (segment) on disk tailored to the requirements of \sys.

\begin{figure}[t!]
  \centering
  % \vspace{-0.4in}
  \includegraphics[width=3in]{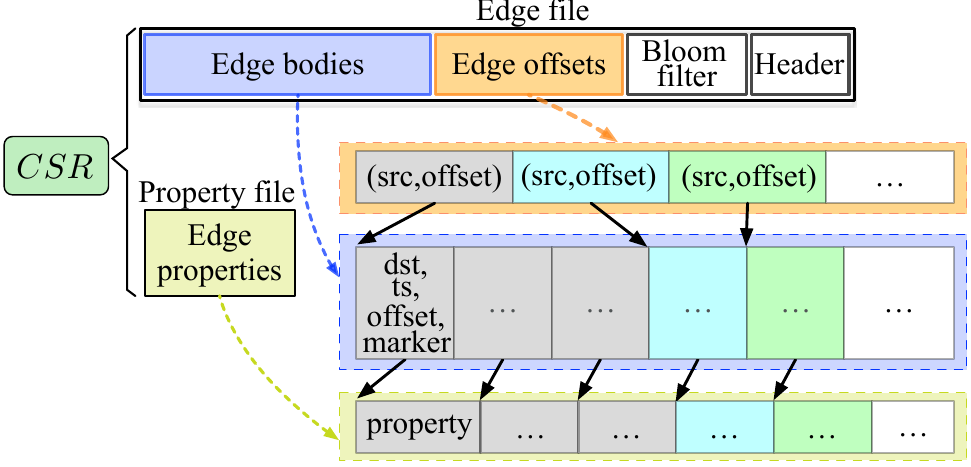}
  \vspace{-0.1in}
  \caption{Details of CSR (segment) file on disk.}
  \label{fig:CSRFiles}
  \vspace{-0.15in}
\end{figure}

\hypertarget{resp:CSRStorageFormat}{  
\stitle{CSR Storage Format}. 
}
Similar to the original CSR, we divide a CSR into two separate files, the edge file and the property file, as shown in Fig. \ref{fig:CSRFiles}. 
For the \textit{edge file}, it comprises four parts, header, Bloom filter, edge offsets and edge bodies.
The header stores the metadata of the edge file, including the number of edge bodies, the number of edge offsets, 
the maximum and minimum source vertex IDs of the edges within the file, and the creation timestamp of the file.
The Bloom filter, a highly compact probabilistic representation of a set, is used to roughly filter whether a particular edge is in the file, which could speed up edge queries.
Specifically, the two vertex IDs (64 bits each) of an edge are hashed into 32-bit integers and then concatenated into a 64-bit integer to serve as the key for the Bloom filter.
Edge offsets record a set of pairs $\langle src, o\!f\!f\!set \rangle$, where
$src$ represents a %records each source 
vertex ID in the current file, and 
$o\!f\!f\!set$ records the offset of the first edge of vertex $src$.
The role of $o\!f\!f\!set$ is similar to the \textit{offset array} in the CSR shown in Fig. \ref{fig:csr}. 
Since each CSR segment only stores edges of a part of vertices,
we use edge offsets instead of the \textit{offset array} in CSR (Fig. \ref{fig:csr}(b)) to reduce index size.
Edge bodies store the data of each edge that comprises
$\langle dst, ts, o\!f\!f\!set, marker \rangle$.
Here, \dst represents the destination vertex ID of the edge, $ts$ denotes the timestamp when the edge was inserted, serving for fine-grained snapshot isolation,
\offset stores the offset of the property of this edge in the property file, as depicted in Fig. \ref{fig:CSRFiles}, and \marker indicates whether the edge is deleted.

Compaction is a key operation to maintain the multi-level structure of \sys, and it determines the data movement between different levels.
We further introduce the details of the compaction in the following, which combines the characteristics of graph data.

\stitle{Vertex-aware Compaction}.  
For the CSR segments that will be compacted, we first select the edges with the smallest source vertex ID from all the CSR segments. 
Then, we sort them in ascending order based on the destination vertex ID, and write them into the same CSR segment sequentially.
If the CSR segment has become very large, we write them into a new CSR segment to keep each CSR segment in a reasonable size. After that, we compact the next vertex's edges similarly. In this way, the edges of each vertex are compacted into the same segment, and the edges are sorted in ascending order based on the destination vertex ID.

\begin{figure}[tbp]
  \centering
  % \vspace{-0.4in}
  \includegraphics[width=3.0in]{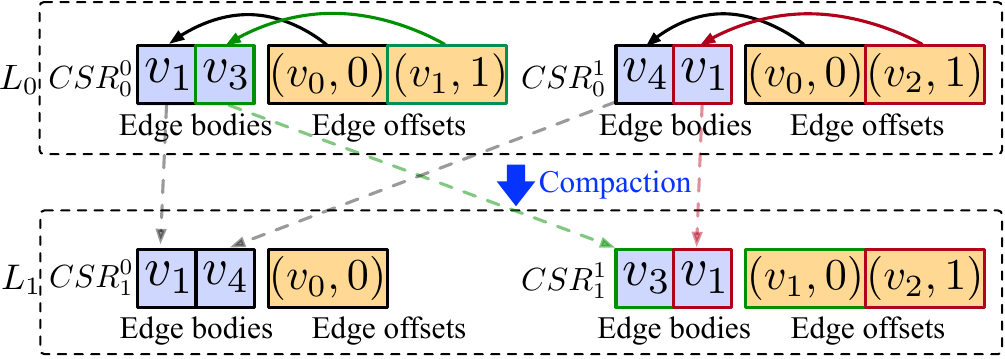}
  \vspace{-0.1in}
  \caption{
  % Compaction of CSR segment.
  An example of compacting two CSR segments.
  }
  \label{fig:compaction}
  \vspace{-0.15in}
\end{figure}

\begin{example}
  Fig. \ref{fig:compaction} 
  shows an example of two CSR segments being compacted from level $L_0$ to level $L_1$, where level $L_1$ is initially empty.
  $CSR^{j}_{i}$ denotes the $j$-th CSR or CSR segment at $L_i$.
  First, consider edge offset ($v_0$, 0) of $CSR^{0}_{0}$ indicating the first edge as edge ($v_0$, $v_1$) and edge offsets ($v_0$, 0) of $CSR^{1}_{0}$ indicating the first edge as edge ($v_0$, $v_4$).
  Obviously, through the offset information, we can determine that they belong to the same source vertex ($v_0$), and the edge ($v_0$, $v_1$) of $CSR^{0}_{0}$  has a smaller destination vertex ID, so it needs to be written to the CSR file ($CSR^{0}_{1}$) of $L_1$.
  Then, consider the next edge offset $(v_{1}, 1)$ of $CSR^{0}_{0}$ indicating $e(v_{1}, v_{3} )$ of a new source vertex ($v_{1}$).
  The edge waiting for compaction in $CSR^{1}_{0}$ is still $e(v_{0}, v_{4})$ of the source vertex $v_0$.
  Therefore, by comparing the source vertices, the source vertex $v_0$ of $CSR^{1}_{0}$ is the unique minimum, and $e(v_{0}, v_{4})$ that will be written to $CSR^ {0}_{1}$.
  Repeat the above process. When the third edge $e(v_{1}, v_{3})$ needs to be written, $CSR^{0}_{1}$ reaches the size limit (if each CSR file is limited to 2 edges), and since this edge is the new source vertex ($v_1$), so this edge is written into the new file $CSR^{1}_{1}$. The final compaction result is shown in Fig. \ref{fig:compaction}.
\end{example}

%%%%%%%%%%%%%%%%%%%%%%%%%%%%%%%%%%%%%%%%%%%%%%%%
%
%   Efficient Multi-level Index
%
%%%%%%%%%%%%%%%%%%%%%%%%%%%%%%%%%%%%%%%%%%%%%%%%

\subsubsection{Efficient Multi-level Index}
\label{sec:disk:index}

Although multi-level CSR combines the excellent write performance of LSM-trees and the efficient read performance of CSR, the multi-level structure of LSM-trees makes reading complex. Since the edges of a vertex may be distributed across multiple levels, existing LSM-tree designs typically use binary search to find the blocks containing the data and then use Bloom filters to determine whether to load those blocks. As analyzed in Section \ref{sec:bg:csrLSM}, this requires numerous random lookups in memory, leading to poor read performance. Therefore, it is essential to design an index to accelerate the location of a vertex's edges in the multi-level structure.

\begin{figure}[tbp]
  \centering
  % \vspace{-0.4in}
  \includegraphics[width=3.3in]{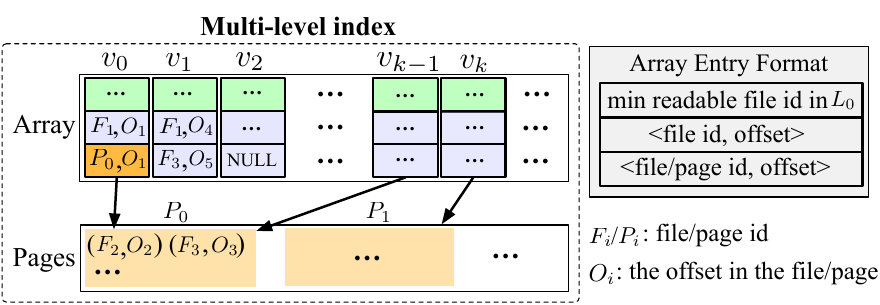}
  \vspace{-0.1in}
  \caption{
  The structure of multi-level index.
  }
  \label{fig:CSROffsetFiles}
  \vspace{-0.15in}
\end{figure}

To efficiently select the CSR segments that contain the edges of the queried vertex, we design a multi-level index that directly locates the CSR segments without numerous random lookups.
Since the edges are stored continuously within each CSR segment, it is sufficient for our index to locate the first edge of each vertex.
In addition, at each level except $L_0$, the edges of each vertex are stored in only one CSR segment.
Therefore, for each level, we can store one position information for each vertex in the index, \ie the file ID of the CSR segment and the offset of the first edge of each vertex. 
A simple method to store the position information of each vertex is to employ an array, which is similar to the offset array in CSR (as shown in Fig. \ref{fig:csr}(b)).
For multi-level CSR, we need to create an array for each level.
However, we found that not every vertex's edges are present at each level, resulting in the arrays always being sparse. 
To compress these sparse arrays and save space, we use the structure shown in Fig. \ref{fig:CSROffsetFiles} to store the position information of vertices.

\stitle{Multi-level Index}.
As shown in Fig. \ref{fig:CSROffsetFiles}, there is an array to store the position information of each vertex, where each entity in the array contains three data items. 
The first item stores a file ID, which points to the first CSR file at $L_0$ that contains the edges of the vertex. Because the edges of one vertex at $L_0$ may spread over multiple CSRs, recording the first CSR file that contains edges can help filter some invalid random reads. 
The second and third data items in the array are used to store the position information of the vertex's edges at different levels, \ie the file ID of the CSR (segment) containing the edge and offset of the first edge in the file. 
We only record two location information in the array based on the following observation. 
When each level is full,
the last two levels at the bottom of multi-level CSR will hold 99\% of the edges because the capacity of each level is tenfold the previous level.
When the number of edge positions for a vertex exceeds 2, \ie the edges of the vertex appear on more than 2 levels, we only use the second data item to store one of the positions, while storing the other position information in a page in the page set (\ie Pages in Fig. \ref{fig:CSROffsetFiles}).
The third data item records the page ID and offset of the position within the page. 
Each page has a size of 4K and is evenly allocated to a continuous interval of size $k$ vertices (default is 1024) during system startup to maintain the locality of index data access. 
Each page records position information, as well as the current storage size and remaining space (not shown in Fig. \ref{fig:CSROffsetFiles}). 
If a page cannot accommodate the indexes of the current interval, the vertex interval is split in half, and the data on the page is written to two new pages, while modifying the recorded page ID and offset in the array according to the vertex interval.
When the data stored in a page becomes too small, it will be considered to be merged with pages of adjacent vertices ranges.
In addition, the multi-level index is synchronously updated during compaction. Since the position information (\ie file ID and offset) of all vertices mirrors the edge offset portion of the CSR file (as shown in Fig. \ref{fig:CSRFiles}), obtaining them is straightforward and does not require additional complex calculations.

Note that the multi-level index is updated in place based on the following three intuitions.
First, the multi-level index is frequently used in query operations and is usually cached in memory, meaning that most update operations can be completed in memory. 
Second, the updated multi-level index does not need to be periodically synchronized to disk as long as the memory is enough. 
Even if it is lost due to a system crash, we can still reconstruct it by the edge offset information from the CSR files on disk. 
Third, the index updates exhibit good spatial locality because each compaction is limited to a range of vertices. 
Thus, even if the indexes are on disk, loading them can still take advantage of the amortized I/O.

By utilizing the multi-level index, \sys can quickly obtain the distribution information of all edges of a vertex at each level by using the vertex ID, avoiding many random lookups.

\begin{example}
Fig. \ref{fig:CSROffsetFiles} presents a simple example of the multi-level index. 
The edges of vertex $v_0$ are stored in three files (\ie $F_1$, $F_2$ and $F_3$) at different levels. 
Therefore, in the multi-level index, only the position information of $v_0$ in the first file is recorded in the array, while the remaining position information is stored on a page with ID $P_0$, and the page ID and offset are recorded in the array. 
For $v_1$, its edge is only stored in files $F_1$ and $F_3$, so its position information is recorded in the array.
\end{example}

\stitle{I/O Analysis}.
%%% write
In terms of write performance, \sys achieves similar write performance to LSM-trees due to its multi-level structure and significantly outperforms CSR.
Specifically, inserting an update into the \mgs entails O(log($d$)) memory I/Os, where $d$ is the average degree of vertices. 
Disk I/O from compaction is similar to LSM-trees, the amortized disk I/O cost for one update is $O(\frac{L \cdot T}{B})$ I/O. 
In addition, during the compaction, one memory I/O is required to update the index of the vertex in the multi-level index.
Notably, if the multi-level index cannot be cached in memory, it will incur at most one disk I/O because compaction usually targets vertices within a contiguous range, 
providing good locality and allowing multiple vertices to amortize the I/O.
%%% read
In terms of read performance, \sys uses a multi-level index and CSR format, resulting in better read performance than LSM-trees. However, due to the multi-level structure, its performance is inferior to that of the original CSR. 
Specifically, locating edges of a vertex in \mgs requires only $O(1)$ memory I/O. Subsequently, through the multi-level index, the position of a vertex's edges in the multi-level CSR can be obtained with 1 or 2 memory I/Os. Note that if the index is on disk, the I/O becomes disk I/O.
In the worst-case workload, where edges are distributed across all levels, loading the edges requires $O(L)$ disk I/O. If properties of edges are also needed, an additional $O(L)$ I/O is required. 
Notably, when executing graph analysis algorithms, \sys not only reduces lookups in memory but also benefits from the graph-aware CSR storage format. 
In comparison, LSM-trees need to parse key-value pairs one by one. We also list the times of memory and disk I/O of \sys in terms of write and read in Table \ref{tab:ReadWriteAnalysis}.

%%%%%%%%%%%%%%%%%%%%%%%%%%%%%%%%%%%%%%%%%%%%%%%%
%
%   Concurrency Control
%
%%%%%%%%%%%%%%%%%%%%%%%%%%%%%%%%%%%%%%%%%%%%%%%%

% \stitle{Read Flow}

\subsection{Vertex-grained Version Control}
\label{sec:design:version}

Consider a concurrent scenario, when reading the graph data in the LSM-tree, the data at $L_i$ is compacting with the data on the $L_{i+1}$, or the \mgs is being flushed to $L_0$. 
After flushing or compacting, the original data, \ie \mgs or CSR (segment) files at $L_i$ will be deleted. 
However, if the original data is deleted before the data is read, it will cause a read error.
Therefore, it is critical to design a version management strategy for the system's files to ensure that read and write tasks run correctly.

Existing LSM-tree based systems \cite{rocksdb,leveldbPaper} maintain a version chain to manage different versions of files. 
Whenever a new file is generated or deleted, a new version is created. 
Each version in the version chain records a readable data set in the system.
The version in the version chain is a static view of all data, which requires that the recorded data cannot change.
However, in \sys, a query task needs to read the multi-level index that needs to be updated during compaction. 
If an immutable copy is generated for the multi-level index and added to the version each time, the cost is intolerable due to frequent compaction.
In \sys, we propose the vertex-grained version control to solve the above problem.

\hypertarget{resp:vertexGrainedVersion}{
\stitle{Vertex-grained Version Control}. 
\sys manages versions in two parts based on the location of the data distribution.
For the data on MemGraph and $L_0$, there is no index to record specific location information (\eg offsets) for each vertex, so the granularity of the version is the entire \mgs and CSR at $L_0$. 
For the data at $L_1$ and subsequent levels, we maintain the index in the multi-level index for each vertex, 
which specifies the data range that each vertex can read, providing version control at vertex granularity.
}

\begin{figure}[tbp]
% \vspace{-0.4in}
  \centering
  \includegraphics[width=3.4in]{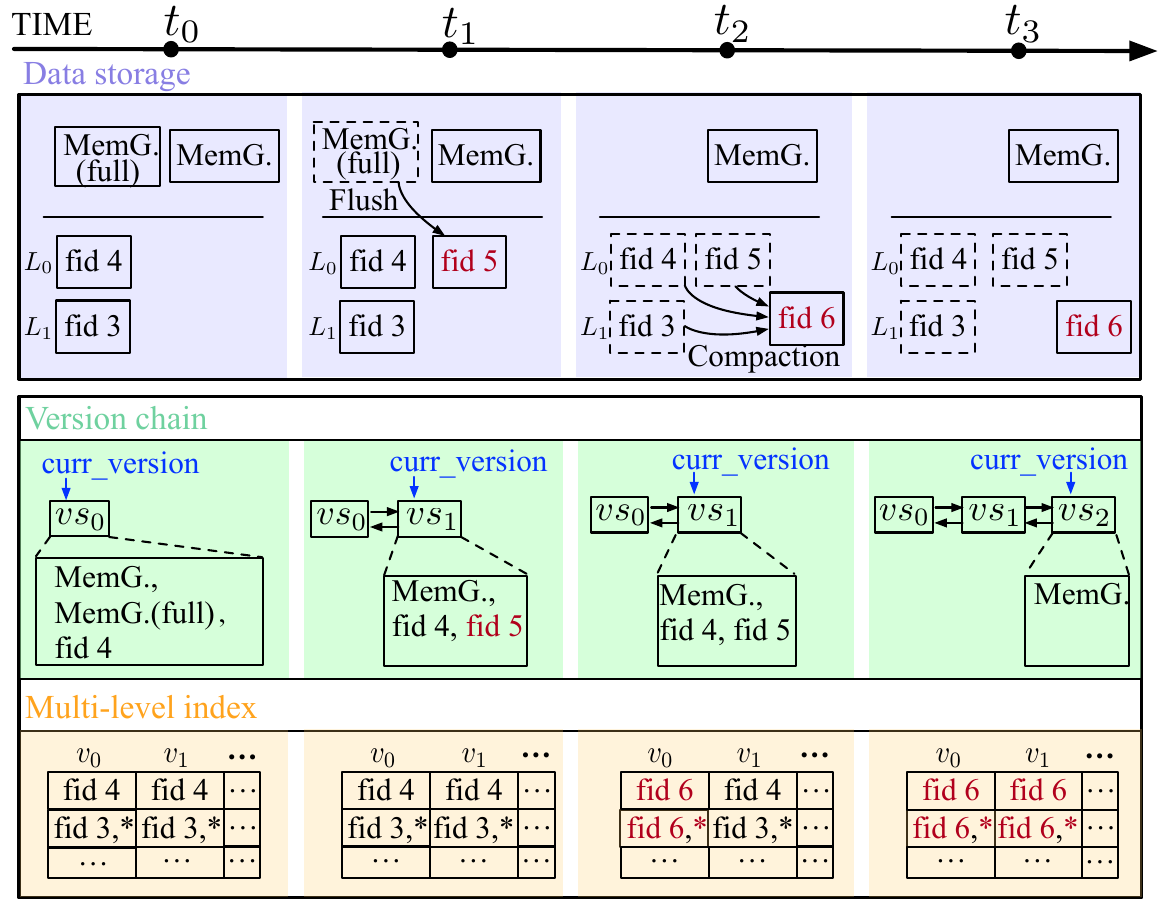}
  % \vspace{0.05in}
  \vspace{-0.1in}
  \caption{
  % Version updates over time.
  An example of a version chain and multi-level index that are updated at different times in response to changes in data storage.
  }
  \label{fig:version}
  \vspace{-0.15in}
\end{figure}

\etitle{Version Control on \mgs and $L_0$}. 
\sys maintains a version chain in memory, with a current version pointer (\ie \textit{curr\_version}) pointing to the most recent version.
Each version records \mgs in memory (including the MemGraph being written and the MemGraph that has reached its capacity, \ie MemG and MemG. full) and the file IDs (\ie fid) of the CSR files at $L_0$.
As shown in Fig. \ref{fig:version}, at time $t_0$, the most recent version is version 0 (\ie $vs_0$) and it records a \mgs (\ie MemG.), a full \mgs (\ie MemG. full), and a CSR file ID (\ie fid 4) at $L_0$.
When a new \mgs is generated, \sys creates a new version by copying the contents of \textit{curr\_version} and adding the new \mgs. 
When a \mgs reaches its capacity, as shown in MemG. full in Fig. \ref{fig:version}, it needs to be flushed to $L_0$.
Once the flush of the \mgs is completed, \sys also generates a new version by copying \textit{curr\_version}, removing the \mgs from it, and adding the file ID obtained from the flush.
As shown in Fig. \ref{fig:version}, at time $t_1$, the MemG. full is flushed to disk and written to the file (\ie fid 5).
At this point, a new version $vs_1$ is generated by copying $vs_0$, removing the MemG. full, and adding the corresponding file ID (\ie fid 5).
Each time a new version is generated, it will be inserted into the version chain as a chain head and become a new \textit{curr\_version}.
The old version may still exist in the version chain, and once no read operation accesses the version, it is automatically removed from the version chain.

\etitle{Version Control at $L_1$ and Subsequent Levels}.
Each compaction causes the data of the old file to be merged and written into the new file.
As shown in Fig. \ref{fig:version}, at time $t_2$, we assume that the vertex ranges of files fid 4 and fid 5 overlap and are selected with file fid 3 at $L_1$ for compaction. They are then written into a new file, fid 6. For simplicity in this example, segmentation of fid 6 is not considered.
Since data on disk is accessed through the multi-level index, we can ensure that the multi-level index points to either the new or the old CSR files, thereby avoiding the reading of duplicate or deleted data.
Specifically, if the compaction occurs at $L_0$, \sys needs to record a minimum readable file ID 
for each vertex involved in the compaction in the multi-level index, \ie $fid+1$, where $fid$ is the maximum file ID at $L_0$ involved in the compaction. 
The minimum readable file ID for $v_i$ indicates that query operations for vertex $v_i$ can only access files at $L_0$ with IDs greater than or equal to this file ID. According to the compaction rules, files at $L_0$ with IDs less than this file ID do not contain edges of $v_i$; otherwise, they would have been compacted to the next level.
In addition, after each compaction is completed, the position information at $L_1$ or subsequent levels recorded in the multi-level index is updated based on the new file ID and the edge offset of each vertex in the new file. 
As shown in Fig. \ref{fig:version}, we illustrate two time points (\ie $t_2$ and $t_3$) during the multi-level index update process. At time $t_2$, only the index of $v_0$ has been updated, while at time $t_3$, the index update of the remaining vertices is completed.
Specifically, at time $t_2$, the index (\ie minimum readable file ID) of $v_0$ at $L_0$ is updated to fid 6 (\ie fid 5 + 1).
It means that $v_0$ can only read files with file ID greater than or equal to fid 6 at $L_0$.
At the same time, the index of $v_0$ at $L_1$ is updated to fid 6 and the new offset (denoted as * in Fig. \ref{fig:version}) in the file.
The updates to the vertex indices are managed using vertex-grained read-write locks to handle read and write conflicts with the reading threads.

Overall, for data on \mgs and $L_0$, \sys uses version chain and $curr\_version$ for version management.
The version chain only maintains the data on \mgs and $L_0$ in \sys, unlike traditional version chains (\eg in RocksDB \cite{rocksdb} and LevelDB \cite{leveldbPaper}), which avoids the overhead of maintaining other levels.
For data at $L_1$ and subsequent levels, \sys uses the multi-level index for version management.
Note that there is some overlap between these two types of data, as data at $L_0$ is written to $L_1$ during compaction. Therefore, \sys ensures global version consistency by cleverly recording the minimum readable file ID in the multi-level index, preventing read operations from accessing duplicate data.

\begin{example}
We use times $t_2$ and $t_3$ in Fig. \ref{fig:version} to explain how to ensure that each vertex reads the correct data during the multi-level index update process.
At time $t_2$, the multi-level index is being updated. 
For example, information for $v_0$ has already been modified, while $v_1$, although needing an update, has not yet started. 
Using \textit{curr\_version} and the multi-level index, $v_0$ can only read \mgs and fid 6. 
Note that files fid 4 and fid 5 will not be read for $v_0$, even though they are in \textit{curr\_version} at $L_0$, because $v_0$ can only read files with IDs greater than or equal to its minimum ID (\ie fid 6) recorded in the multi-level index.
For $v_1$, the data retrieved includes \mgs, fid 4, fid 5, and fid 3. 
% Despite $v_0$ and $v_1$ reading different files, their read content is equivalent due to the compaction process not altering the data content,\ie the contents of file fid 6 are equivalent to files fid 4, fid 5, and fid 3.
Although $v_0$ and $v_1$ read different files, 
the compaction process merely merges the data, ensuring that the data they access is equivalent; specifically, the data in file fid 6 is identical to that in files fid 4, fid 5, and fid 3.
At $t_3$, compaction is complete, information for $v_1$ has already been updated in the multi-level index, and $vs_2$ is created by deleting fid 4 and fid 5 from $vs_1$.
% All vertices are read by $curr\_version$ and the multi-level index.
\end{example}

Vertex-grained version control ensures each vertex can access a unique data view during flush and compaction. However, concurrent read-write scenarios still require concurrency control strategies to provide correct data access among multiple read and write threads.

\hypertarget{resp:ConcurrentReadWrite}{
\stitle{Concurrent Read and Write}}.
There are two kinds of conflicts in our \sys
to handle, write-write and read-write conflicts.

\etitle{Write-write Conflict}. 
As shown in Fig. \ref{fig:overview}, a write-write conflict only occurs in \mgs because updates are always written to \mgs first in \sys. 
In \mgs, each vertex's edge is written to its independent array position or skip list, so there are no conflicts between different vertices. When multiple write threads attempt to insert edges for the same vertex simultaneously, \sys uses vertex-grained write locks to ensure safety. They must acquire the write lock for the vertex before performing the write operation. This is a common approach that many existing dynamic graph storage systems have adopted \cite{livegraph, sortledton, gart, llama, terrace}.

\etitle{Read-write Conflict}. 
It occurs when a read operation accesses a vertex's index from the multi-level index while a compaction thread needs to modify the index.
\sys ensures read-write safety through vertex-grained read-write locks.
A vertex's read-write lock allows multiple read threads to simultaneously acquire the read lock to access the vertex's index, while only one write thread can acquire the write lock to modify the index. This ensures both read-write safety and high performance.
Then, based on the obtained index and \textit{curr\_version}, we can further access the specific data.
When the data being read is in \mgs, the read operation does not need to acquire any locks. 
This is because the read operation acquires a timestamp $\tau$ before reading, and it only reads edge with timestamps less than or equal to $\tau$. Thus, even if new data is inserted during the reading process, it will not be read because its timestamp must be greater than $\tau$. When the data being read is in the CSR or CSR segment files, there are no read-write conflicts because these files are not modified.

In concurrent read-write scenarios, it is critical to ensure that read tasks access consistent graphs at different times.
For example, to achieve correct results in graph analysis, it is necessary to perform each iteration on a consistent graph, \ie on the same snapshot. The following paragraph introduces how \sys implements this.

\hypertarget{resp:ReadGraph}{
\stitle{Read Graph with Vertex-grained Version Control}}.
\sys provides a consistent graph for a long-running query (\eg graph analysis) by using timestamps recorded on the edges as snapshots. 
Specifically, as shown in Fig. \ref{fig:CSRFiles}, each edge records a timestamp (\ie $ts$) when it is inserted, representing a snapshot. 
When a graph analysis task is executed, it first acquires the current latest snapshot number (\ie timestamp), denoted as $\tau$. 
% Subsequently, each time \sys reads edges of a vertex,
Subsequently, each time \sys reads the edges of a vertex, it first uses vertex-grained version control to retrieve readable data range (including \mgs and CSRs on disk) from the multi-level index and the latest version.
% \sys only reads edges with timestamps less than or equal to $\tau$. 
\sys reads data within the range and only reads edges with timestamps less than or equal to $\tau$. Additionally, any edge with a deletion marker is discarded.
If new edges are continuously inserted during the graph analysis process, these new edges will have timestamps greater than $\tau$ and will not be read by the current graph analysis task, ensuring that the analysis is always performed on the same snapshot ($\tau$).

%% file: 6-eval.tex
\section{Evaluation}
\label{sec:expr}

\begin{table}[t!] % "h" means "here", "t" means "top"
  \caption{Graph datasets used in the experiments.}
  \label{tab:dataset}
  % \hspace{-0.25in}
  % \vspace{-0.15in}
  \centering
  \small
  % \footnotesize
  % {\renewcommand{\arraystretch}{1.2}
  \setlength{\tabcolsep}{3pt} %colums
  \begin{tabular}{cccccc}
  \toprule
   Dataset & $|V|$ & $|E|$ & Avg. Deg. & Size & Type \\
  \midrule
  IT-2004 \cite{it-2004} & 41,291,594 & 1,150,725,436  & 28 & 18GB & \multicolumn{1}{c}{\multirow{2}*{Web Graph}} \\
  UK-2007 \cite{uk-2007} & 105,153,953 & 3,301,876,564 & 31 & 50GB \\
  Twitter \cite{twitter} & 41,652,230 & 1,468,365,182 & 35 & 22GB & \multicolumn{1}{c}{\multirow{2}*{Social network}} \\
  Friendster \cite{friend} & 68,349,467 & 2,586,147,869 & 38 & 39GB \\
  \bottomrule
  \end{tabular}
  \vspace{-0.15in}
\end{table}

\newcommand{\tw}{0.24\textwidth} 
\newcommand{\hsp}{\hspace{1mm}}

\subsection{Evaluation Setup}
\label{sec:expr:setup}

\hypertarget{resp:implementation}{
\stitle{Implementation}.
}
\sys is implemented in approximately 13,000 lines of C++ code, including the support for various analytical algorithms and the validation for the proposed functionalities. It supports both plain graphs and property graphs with 4-byte or 8-byte vertex sizes (8 bytes in our default setting).
The capacity of \mgs is limited to 64MB,
and the capacity of each level on the disk grows by a factor of 10, with a maximum of 5 levels. Additionally, two \mgs are allowed to alternately accept updates in memory.
For various types of properties (\eg weights), we uniformly store them as string types on the edges. 
\sys primarily focuses on the design and optimization of edge storage, and we adopt a similar approach to LiveGraph \cite{livegraph} for storing vertices by appending them to a vertex file.
Additionally, we use a vertex ID recycling strategy similar to that in LiveGraph \cite{livegraph} and RisGraph \cite{risGraph} to manage deleted vertices, \ie the IDs of deleted vertices are reused by recycling the IDs of the deleted vertices and assigning them to the newly inserted vertices.
% \sys supports the storage of both outgoing and incoming edges. 
% By default, the system only stores the outgoing edges of the graph as it is sufficient to meet the requirements of many graph analysis algorithms. 

\stitle{Evaluation Platform}.
Experiments are conducted on AliCloud ecs.i2.4xlarge instance, which contains 16 hyperthread vCPU cores, 128GB memory (33MB L3 Cache), and 2 SSDs of 1.7TB, which can achieve up to 500MBps read/write sequential performance.
All experiments limit their memory to 16GB using the Linux cgroup tool \cite{livegraph}.
The instance runs Ubuntu 18.04 with Linux kernel version 4.15.0-173-generic.
All codes are compiled using GCC v11.4.0.

\stitle{Graph Algorithms}.
We consider four typical graph analysis algorithms in our experiments, including Single Source Shortest Path (SSSP), Breadth-First Search (BFS), Connected Component (CC), and SCAN.
SCAN refers to traversing all one-hop neighbors of each vertex, which is a fundamental operation in many graph algorithms, such as PageRank, PHP, and GNN.
We use the same implementations of all graph algorithms to ensure a fair comparison among various storage systems (except for MBFGraph, we use its own test code because it employs the edge-centric computation model).

\stitle{Graph Datasets}.
We adopt four real-world graphs in our experiments, as shown in Table \ref{tab:dataset}, including two web graphs, IT-2004 (IT) \cite{it-2004} and UK-2007 (UK) \cite{uk-2007}, and two social networks, Twitter (TW) \cite{twitter} and Friendster (FS) \cite{friend}.
These datasets are provided as binary files with 8-byte vertex IDs.
We randomly inserted edges from the datasets into each system in all experiments. When evaluating the insertion performance, we first insert 80\% of the data to form a baseline and then use the remaining data to evaluate the performance~\cite{warmupYCSB,dgap}.
For mixed updates,
we default to a ratio of 20:1 for insertions and deletions \cite{DBLPdel}.

\hypertarget{resp:Competitors}{
\stitle{Competitors}.
}
We compare {\sys} with four state-of-the-art dynamic (graph) storage systems that support updates and analysis on disk, LiveGraph \cite{livegraph}, LLAMA \cite{llama}, MBFGraph \cite{mbfgraph} and RocksDB \cite{rocksdb}. 
LiveGraph \cite{livegraph} is an advanced dynamic storage system mainly designed for memory scenarios. It also uses \mmap to support scenarios where the data volume exceeds memory capacity. 
It stores each vertex's neighbors contiguously to improve the performance of reading the edges of a vertex.
LLAMA \cite{llama} stores graphs as a time series of snapshots, where each snapshot is similar to a CSR. 
An excessive number of snapshots can lead to performance degradation.
LLAMA creates a new snapshot every 10 seconds as recommended by the author.
However, under this setting, LLAMA does not work on the dataset we used.
Therefore, we adjusted the interval to 40 seconds.
MBFGraph \cite{mbfgraph} is a state-of-the-art SSD-based analytics system for evolving graphs. 
It achieves ultimate update performance by directly appending updates to the end of the file. 
However, this design makes MBFGraph only suitable for the edge-centric
computation model.
% Rocksdb 
RocksDB~\cite{rocksdb} is a well-known and widely used open-source key-value store based on LSM-tree. 
In RocksDB, the adjacency list is represented as a single sorted collection of edges, whose unique key is a vertex ID pair (\ie $\langle \text{src, dest} \rangle$) \cite{livegraph}.
To be fair, we turn off WAL (Write-Ahead Log) in Rocksdb because other systems do not support it.
In addition, for other parameters, we adopt RocksDB's default configuration.
All systems use 16 threads by default, with 8 threads allocated for reading and 8 threads for writing in mixed read-write workloads. For write tasks, RocksDB and LSMGraph use half of the threads for background compaction.

\begin{figure}[tbp]
  % \vspace{-0.41in}
  \centering
  \begin{subfigure}[b]{0.25\linewidth}
      \includegraphics[width=\textwidth]{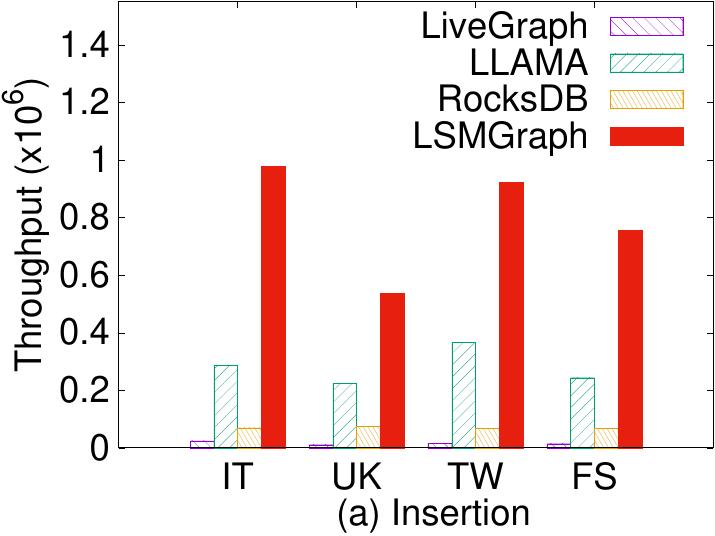}
      % \caption{Insertion}
      \label{fig:throughput:add}
  \end{subfigure}
  \ \ \
  %\hspace{0.1in}
  \begin{subfigure}[b]{0.25\linewidth}
      \includegraphics[width=\textwidth]{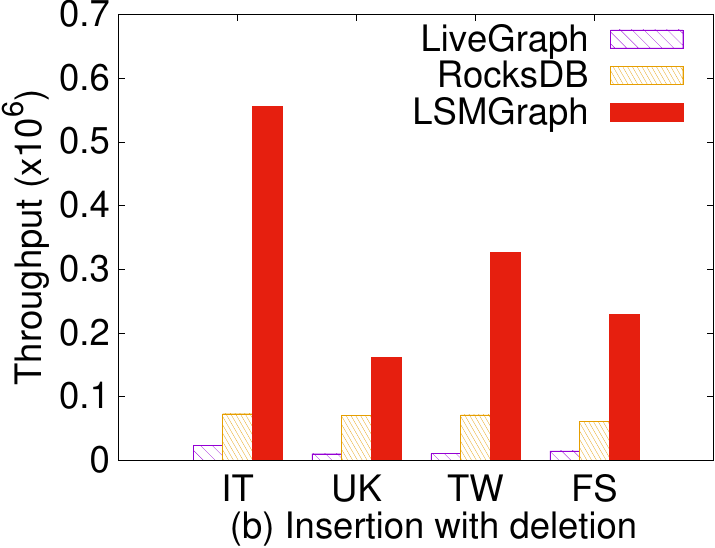}
      % \caption{Insertion with deletion}
      \label{fig:throughput:del}
  \end{subfigure}
  
  \vspace{-0.2in}
  \caption{Throughput comparison of graph updates.}
  \label{fig:throughput}
\end{figure}

\subsection{Graph Update Performance}
\label{sec:expr:insert}

We first evaluate the performance of different systems for ingesting graph updates on four real-world datasets, and the results are shown in Fig. \ref{fig:throughput}.
Fig. \ref{fig:throughput}(a) shows the throughput (edges per second) of all systems when performing only edge insertions. Fig. \ref{fig:throughput}(b) shows the throughput of insertions with random delete operations (4.76\% of all the operators).
As mentioned in the documentation of the LLAMA code repository \footnote{https://github.com/goatdb/llama}, LLAMA cannot handle deletion operations correctly. 
Therefore, the results for LLAMA are not reported in Fig. \ref{fig:throughput}(b).
In addition, MBFGraph provides a reference implementation that utilizes the Linux \textit{cat} command and output redirection to append the update file to the original edge file. 
Using this method, we test the throughput of MBFGraph with different datasets, which obtains the average throughput $3\times10^7$, surpassing other systems 30-2400$\times$.
However, we find that the significant improvement primarily arises from the direct file operations performed by MBFGraph, rather than individual edge insertions. 
Since this paper focuses on the real-time data ingestion capability of the systems, we mainly discuss graph update performance differences between \sys and other competitors.

\begin{figure}[tbp]
  % \vspace{-0.41in}
  \centering
  \begin{subfigure}[b]{0.25\linewidth}
      \includegraphics[width=\textwidth]{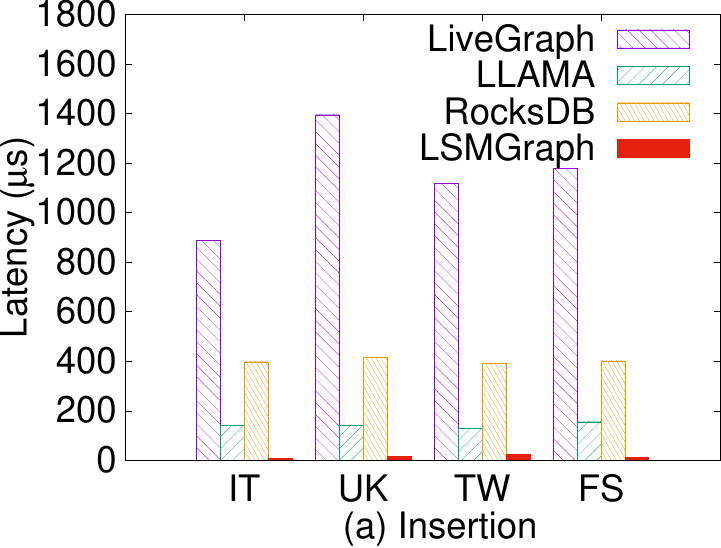}
      % \caption{Insertion}
      \label{fig:p99:add}
  \end{subfigure}
  \ \ \
  %\hspace{0.1in}
  \begin{subfigure}[b]{0.25\linewidth}
      \includegraphics[width=\textwidth]{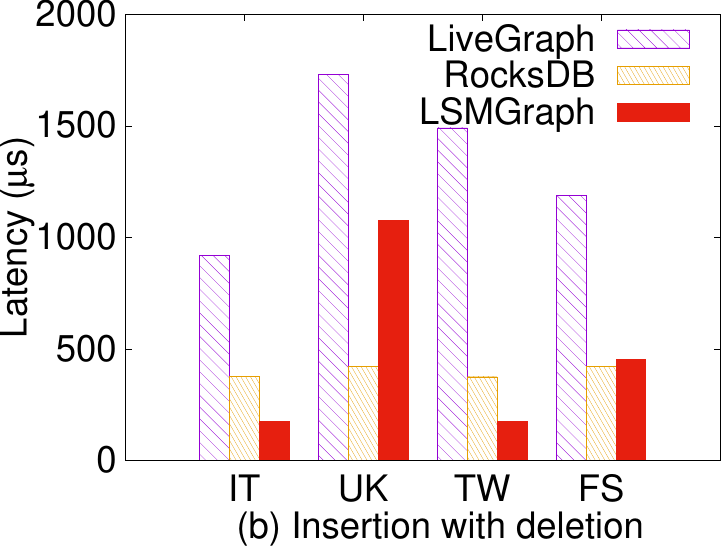}
      % \caption{Insertion with deletion}
      \label{fig:p99:del}
  \end{subfigure}
  
  \vspace{-0.2in}
  \caption{P99 latency comparison of graph updates.}
  \label{fig:p99}
  \vspace{-0.15in}
\end{figure}

As shown in Fig. \ref{fig:throughput}(a), \sys achieves significantly higher throughput than other systems across all datasets. 
Specifically, \sys achieves an average speedup of 50.56$\times$ (up to 58.71$\times$) over LiveGraph.
The main benefit of \sys comes from its adoption of an LSM-tree structure, where updates are first cached in memory and then sequentially written to disk.
In contrast, LiveGraph performs in-place updates for inserted data and constantly adjusts the size of its edge vector to ensure continuous storage of neighbors, resulting in the lowest throughput compared to all other systems.
Compared to LLAMA, \sys achieves an average speedup of 2.85$\times$ (up to 3.40$\times$) over LLAMA.
After \mgs reaches its capacity limit, \sys allows for asynchronous flushing to disk and a new \mgs can continuously accept new updates.
In contrast, LLAMA can only insert memory data and write it to disk sequentially to maintain and update links between different versions.
In addition, since \sys employs an LSM-tree structure, the data cached in memory at one time can be much smaller than in LLAMA, and the conversion to the CSR file format on disk is faster each time it is flushed.
Besides, \sys achieves an average speedup of 11.60$\times$ (up to 14.15$\times$) over RocksDB.
These benefits primarily come from the efficient memory cache structure, \mgs, which is specifically designed for graph data in \sys.
In addition, RocksDB employs a group commit approach before inserting data into the MemTable in memory, where only the group leader can perform the insertion operation, impacting its insertion performance.

Fig. \ref{fig:throughput}(b) demonstrate similar performance trends as Fig. \ref{fig:throughput}(a), and \sys continues to exhibit outstanding performance advantages. It is worth noting that graph systems experience some performance degradation in deletion scenarios. This is because deletion operations require verifying the existence of the edges to be deleted, resulting in additional query operations compared to pure insertion operations.

\hypertarget{resp:UpdatePerformance}{
We also evaluate the p99 latency of different systems during the above ingestion update process. %, as shown in Fig. \ref{fig:p99}.
Fig. \ref{fig:p99}(a) shows the p99 latency of each system with only insert operations;
\sys exhibits significantly lower latency than other storage systems, thanks to \sys's multi-level write-optimized structure and graph-aware cache structure in memory (\ie \mgs), which enable fast updates. 
Fig. \ref{fig:p99}(b) shows the p99 latency with inserts accompanied by random delete operations; \sys still demonstrates lower latency than other systems in most cases. However, these latencies are higher compared to scenarios with only insert operations. This is because, like other graph systems \cite{livegraph,llama,sortledton}, \sys needs to look up an edge before deleting it, causing random disk I/O. 
In contrast, RocksDB treats deletions the same as insertions without performing lookup operations.
Although LiveGraph also performs lookup operations during deletions, its in-place updates cause disk I/O even for insertion operations. Therefore, its latency is consistently high whether deletions are included or not, exhibiting little variation.
}

\subsection{Graph Analysis Performance}
\label{sec:expr:alg}

\begin{figure*} %[tbp]
  \centering
  % \vspace{-0.35in}
  \includegraphics[width=0.95\linewidth]{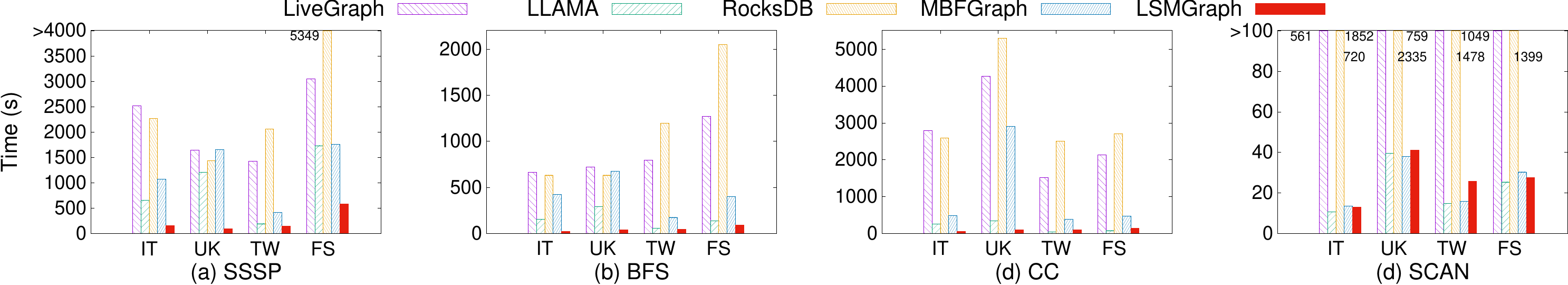}
  \vspace{-0.1in}
  \captionsetup{font=small}
  \caption{Running time comparison of graph analysis.}
  \label{fig:algRumtime}
  % \vspace{-0.09in}
\end{figure*}

\begin{figure*} %[tbp]
  \centering
  \includegraphics[width=0.95\linewidth]{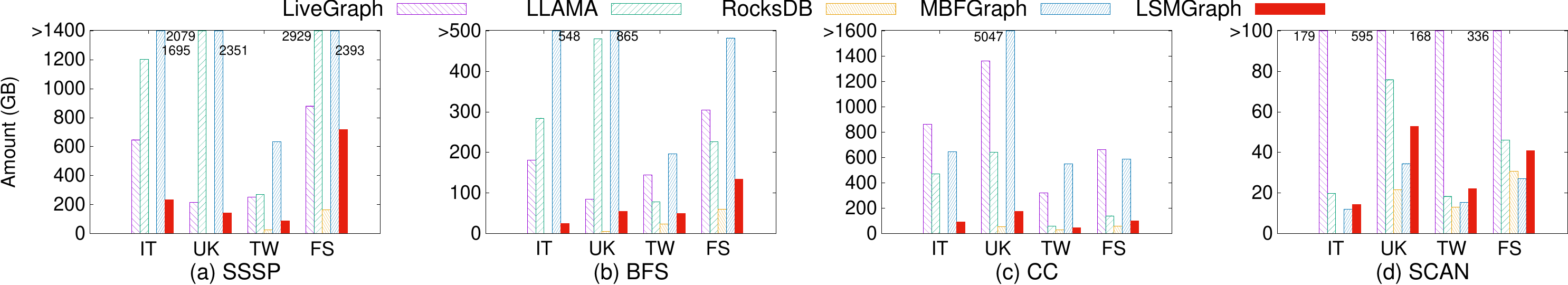}
  \vspace{-0.1in}
  \captionsetup{font=small}
  \caption{I/O amount comparison of graph analysis. }
  \label{fig:algIO}
  % \vspace{-0.25in}
\end{figure*}

We next evaluate the performance of all systems in executing graph analysis on different graph datasets. 
We also measure the I/O amount during the execution of the algorithms (default representation in the experiment indicates disk I/O).
The results are shown in Fig. \ref{fig:algRumtime} and Fig. \ref{fig:algIO}, respectively.
Overall, the results indicate that \sys outperforms other systems in most cases. 
Specifically, \sys achieves an average speedup of
24.4$\times$ (up to 45.1$\times$) over LiveGraph, 
3.1$\times$ (up to 12.1$\times$) over LLAMA, 
30.8$\times$ (up to 57.4$\times$) over RocksDB,
and
6.6$\times$ (up to 27.2$\times$) over MBFGraph.

%%%%%%%%%%%%%%%%%%%%%%%%%%%%%%%%%%%%%%%%%%%%%%%%%%
%%%%% Graph analysis time

%%%%%%%%%%%%%%%%%%%%%%%%%%%%%%%%%%%%%%%%%%%%%%%%%%

%%%%%%%%%%%%%%%%%%%%%%%%%%%%%%%%%%%%%%%%%%%%%%%%%%
%%%%% Graph analysis IO
%%%%%%%%%%%%%%%%%%%%%%%%%%%%%%%%%%%%%%%%%%%%%%%%%%

For SSSP, BFS, and CC, \sys demonstrates significant performance advantages in graph analytics. Compared to LiveGraph, LLAMA, and MBFGraph, \sys exhibits a lower I/O amount and shorter algorithm execution time. 
For LiveGraph, the edges of different vertices are scattered and distributed across the disk, resulting in severe read amplification and ultimately leading to degraded performance. 
LLAMA generates new snapshots continuously to handle updates, requiring the reading of multiple snapshots when accessing all edges of a vertex. However, due to the use of a CSR-like representation for each snapshot, LLAMA achieves better locality between different vertices, resulting in superior algorithm performance compared to other competitors. 
RocksDB, designed for key-value pairs, has the lowest storage size (analysis in Section \ref{sec:expr:spaceCost}), which makes it have the lowest I/O amount. 
However, compared to all other graph systems, it has the worst graph analysis performance due to its complex lookup process and inefficient traversal performance of SSTables.
% For example, our analysis finds that on IT, locating data alone accounted for 36\% of the total time. 
By employing \mgs, multi-level index, and CSR, \sys achieves better lookup and traversal performance than RocksDB. 
MBFGraph has almost the highest I/O amount because it adopts an edge-centric computation model. For these algorithms, even if only a small number of vertices need to be processed, it still needs to traverse the entire edge file, leading to poor algorithm performance.
It is noteworthy that the trend of changes in the amount of I/O and running time are not always completely consistent. For example, in Fig. \ref{fig:algRumtime}/\ref{fig:algIO}(a), the I/O amount for LLAMA and MBFGraph is higher than that for LiveGraph, yet the running time is shorter. 
This is mainly due to LiveGraph's scattered and disordered storage structure, which leads to more random I/Os when retrieving edges for different vertices. 
In contrast, LLAMA and MBFGraph benefit from more sequential I/O operations due to their compact CSR-like storage structure and edge-centric computation model, respectively. 
Particularly, MBFGraph requires sequential scanning of the entire edge file in almost every iteration. 
Although the volume of data read is large, the sequential I/O access pattern better leverages the high data transfer rates of SSDs.

For SCAN, it traversals all one-hop neighbors of each vertex. 
In this sequential traversal mode, LLAMA, MBFGraph, and \sys demonstrate comparable performance and outperform the other two systems. 
Among them, LLAMA and \sys utilize their CSR structures to achieve the greatest spatial locality \eat{advantage }in sequential traversal mode.
Compared with LLAMA, \sys achieves comparable performance while storing more information in each edge to provide fine-grained snapshot isolation support for real-time graph analysis, which is a feature that LLAMA's coarse-grained snapshots do not possess. 
In addition, for MBFGraph, regardless of how much data in the graph is involved in the computation, almost the entire edge file needs to be loaded into memory.
Since SCAN requires the participation of all graph data, MBFGraph performs better in running SCAN compared to the first three algorithms.

\begin{figure*} %[tbp]
%\vspace{-0.3in}
\centering
 % \flushleft
% \hspace{-0.23in}
% \hspace{-0.1in}
    \begin{minipage}{0.3\textwidth}
        % \vspace{0.07in}
        \begin{center}
        \includegraphics[width=0.7\linewidth]{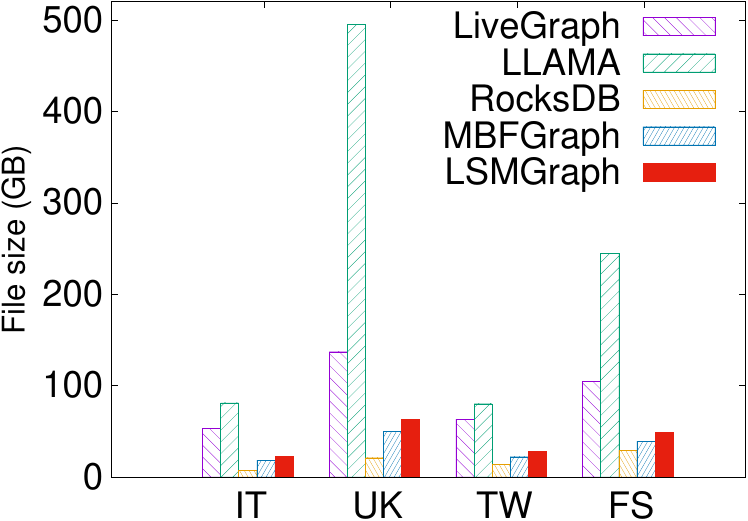}
        \captionsetup{font=small}
        \caption{Space cost comparison.}
        \label{fig:filesize}
        % \vspace{-0.12in}
        \end{center}
    \end{minipage}%
    \ \ \
    \begin{minipage}{0.65\textwidth}
        \vspace{-0.05in}
        \hspace{-0.4in}
        \begin{center}
        \subfloat{\includegraphics[width = 0.33\linewidth]{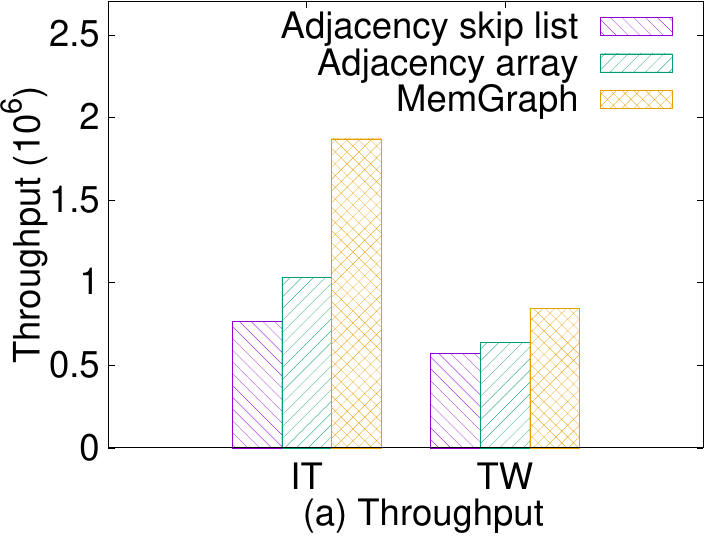}\label{fig:memgraph:throughput}
        }
        \hspace{40pt}
        \subfloat{\includegraphics[width = 0.33\linewidth]{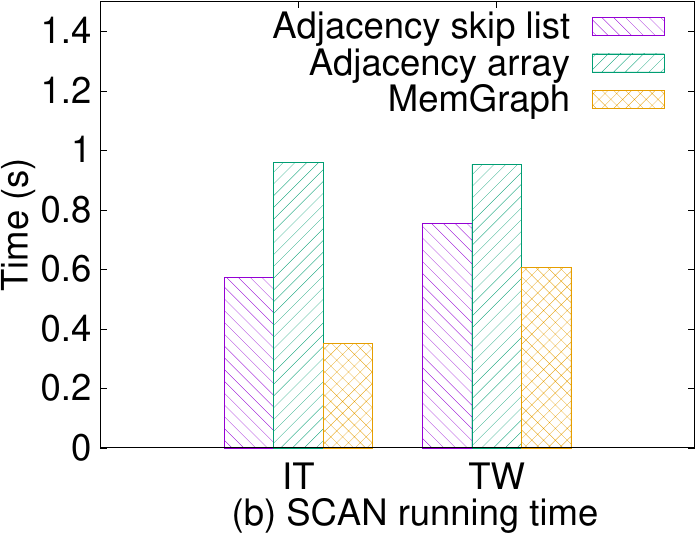}\label{fig:memgraph:scan}}
        \vspace{-0.06in}
        \captionsetup{font=small}
        \caption{Performance comparison of different memory cache structures.}
        \label{fig:memgraph}
        \end{center}
    \end{minipage}
    \hspace{0in}
\end{figure*}
% }

\begin{figure*} %[tbp]
%\vspace{-0.3in}
\centering
    \begin{minipage}{0.42\textwidth}
        \begin{center}
            \subfloat{\includegraphics[width = 0.5\linewidth]
                {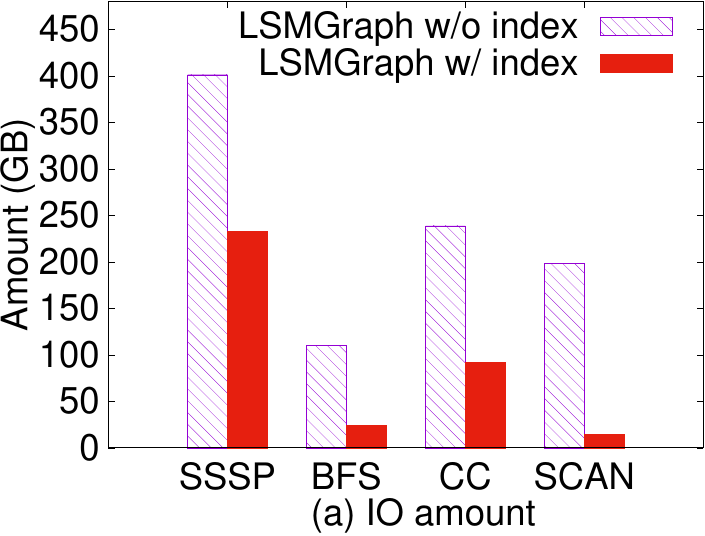}\label{fig:indexIO} \hspace{0in}}
            \subfloat{\includegraphics[width = 0.5\linewidth]
                {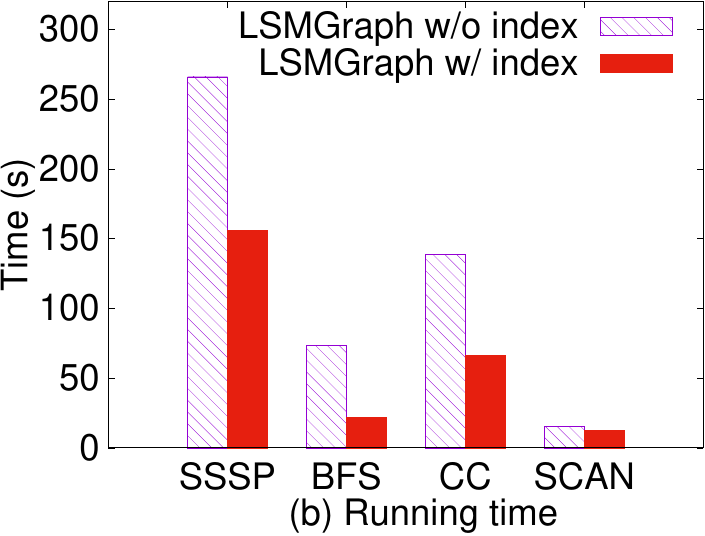}\label{fig:indexTime}}
            % \vspace{-0.12in}
            \captionsetup{font=small}
            \caption{The effectiveness of multi-level index.}
            \label{fig:multiIndex}
        \end{center}
    \end{minipage}
    \ \ \
    \begin{minipage}{0.54\textwidth}
      \centering
      \begin{subfigure}[b]{.38\textwidth}
          \includegraphics[width=\textwidth]{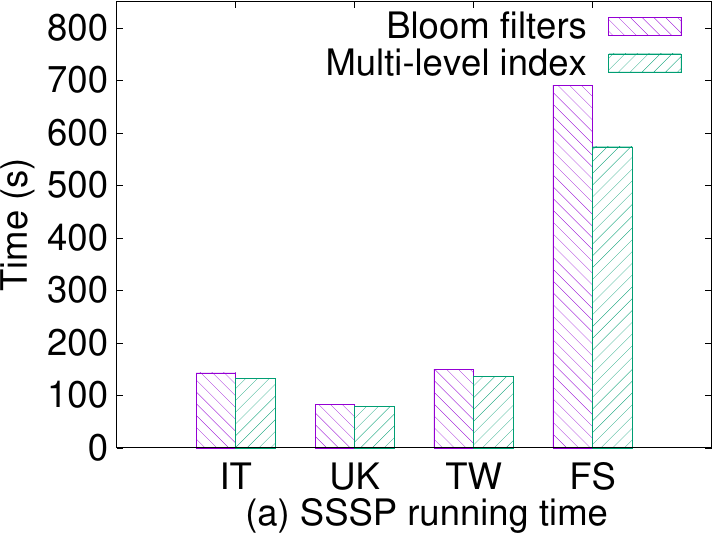}
          \label{fig:bloomfilter:scan}
      \end{subfigure}
      % \hspace{0.19in}
      \begin{subfigure}[b]{.38\textwidth}
          \includegraphics[width=\textwidth]{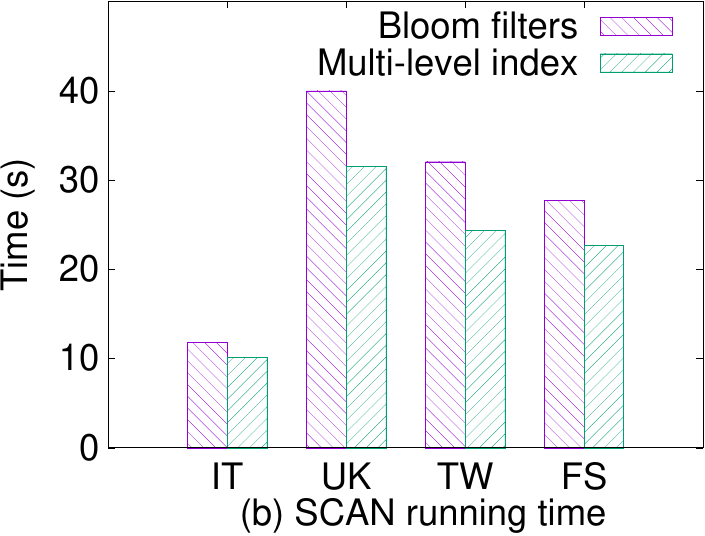}
          \label{fig:rbloomfilter:sssp}
      \end{subfigure}
      \vspace{-0.15in}
      \caption{Comparison of Bloom filters and Multi-level index.}
      \label{fig:bloomfilter}
    \end{minipage}

  % \vspace{-0.15in}
\end{figure*}

\subsection{Space Cost}
\label{sec:expr:spaceCost}

We also evaluate the disk space occupied by \sys and other systems after inserting four real-world datasets.
The results are shown in Fig. \ref{fig:filesize}. 
Compared to LiveGraph and LLAMA, \sys has minimal storage overhead. 
Specifically, the average storage overhead of \sys is only 45\% of LiveGraph and 24\% of LLAMA. 
LiveGraph's larger storage size primarily results from the significant replication of edge blocks and its inefficient recycling mechanism. LLAMA's larger storage size stems from the substantial data redundancy between multiple snapshots.
Compared to MBFGraph, \sys consumes 27\% more space on average.
This is because MBFGraph only stores the topology and properties of the graph, while \sys, as well as LiveGraph, stores timestamps additionally to achieve snapshot isolation.
It is worth noting that RocksDB has the lowest space overhead compared to all native graph systems. 
That's because RocksDB stores vertex IDs as strings, which can save a significant amount of space when vertex IDs are small, and it uses compression algorithms to further compress space, which is orthogonal to storage structure design and can also be used in other graph storage systems, including \sys.
However, these methods lead to higher parsing and comparison costs during the lookup and traversal processes.

\subsection{Effectiveness of Memory Cache Structure}

To evaluate the design of \mgs, we compare the throughput of graph updates and the scan time when using only the adjacency array and only the adjacency skip list, as shown in Fig. \ref{fig:memgraph}. 
The skip list implementation is sourced from RocksDB \cite{rocksdb}, which is a lock-free and efficient implementation that supports concurrent read and write operations. 
The adjacency array is our implementation, which ensures concurrent read and write operations using read-write locks. 
\sys outperforms the comparative structures in terms of throughput and scan time on both datasets. 
For read performance, \sys improves spatial locality by storing edges of low-degree vertices in the single large edge array, thereby avoiding the need to scan multiple small arrays or multiple skip lists, which enhances read performance.
For write performance, the high flexibility of skip lists is hard to show with low data volumes and can disrupt the spatial locality of the data.
Therefore, when inserting edges of low-degree vertices, \sys uses a common array of edges to avoid issues with frequent small memory allocations and expansions.
For high-degree vertices, \sys employs skip lists to increase flexibility, avoiding data migration when inserting edges into the array.

\subsection{Effectiveness of Multi-level Index}
\label{sec:expr:levelindex}

To verify the effectiveness of the multi-level index on system performance, we compare the performance of \sys with the multi-level index (\sys w/ index) and \sys without the multi-level index (\sys w/o index) 
on different graph analysis algorithms. 
When \sys does not use the multi-level index and instead adopts a naive index approach, it utilizes range indexing on the file level and offsets information within the CSR file for lookup. This method is similar to the indexing approach used in LSM-tree implementations like RocksDB.
We measure the running time and I/O amount generated by the two methods on IT, and the results are shown in Fig. \ref{fig:multiIndex}.
It can be observed that adopting the multi-level index effectively reduces the I/O amount and the running time of various graph analysis algorithms. 
These benefits mainly stem from the fact that the multi-level index almost only requires a single random I/O to obtain the distribution information of a vertex's edges on each level. 
It can also quickly determine whether edges exist on a specific level, avoiding unnecessary read operations. 
Therefore, the reduction of the I/O amount here is the key to performance gain.

We also compare the efficiency of Bloom filters in RocksDB and \sys's multi-level index on graph analysis performance. 
Fig. \ref{fig:bloomfilter} shows the running time of SSSP and SCAN on four datasets using the two different indexing schemes. 
It can be seen that the multi-level index is more efficient than Bloom filters in these two classical graph workloads. 
Specifically, the multi-level index outperforms the Bloom filter by an average of about 1.05-1.32$\times$.
This gain comes from the fact that multi-level indexes always reduce random memory accesses.

\begin{figure}
  % \hsp
  % \vspace{-0.3in}
  \begin{subfigure}[b]{.25\textwidth}
      \includegraphics[width=\textwidth]{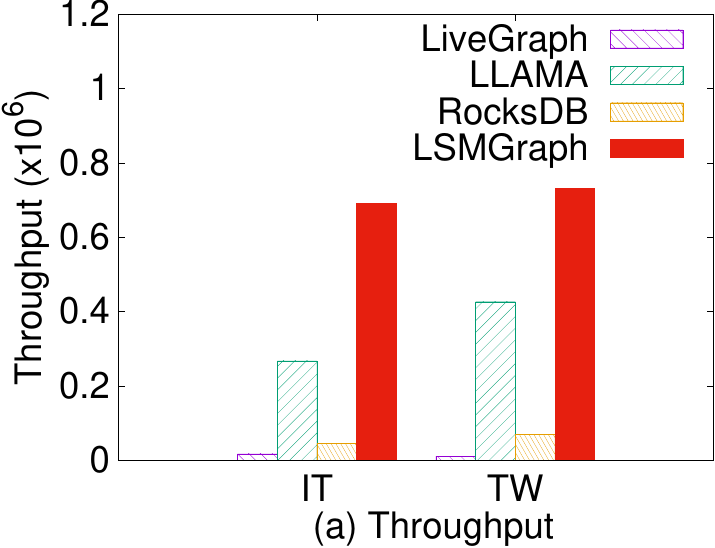}
      \label{fig:rw:throughput}
  \end{subfigure}
  % \hsp
  \hspace{0.19in}
  \begin{subfigure}[b]{.25\textwidth}
      \includegraphics[width=\textwidth]{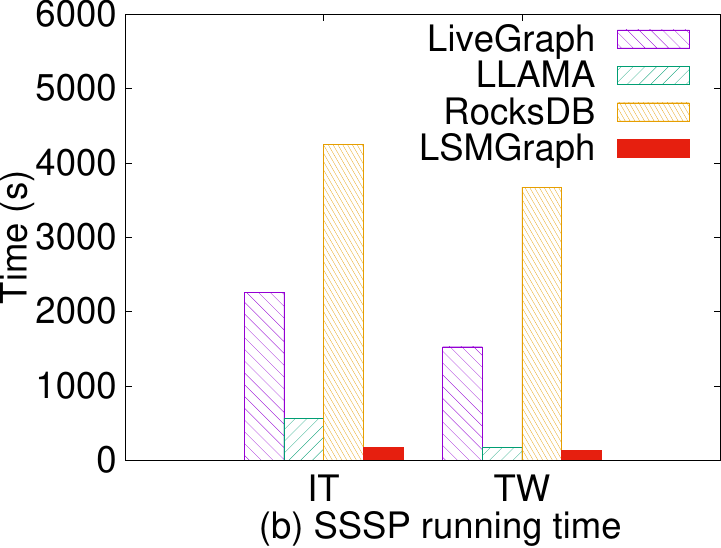}
      \label{fig:rw:runtime}
  \end{subfigure}
  \vspace{-0.2in}
  \caption{Performance comparison under update-analysis mixed workload.}
  \label{fig:rw}
  \vspace{-0.15in}
\end{figure}

\subsection{
Performance under Update-analysis Mixed Workload
}
\label{sec:expr:mixworkload}

We analyze the performance of \sys in executing concurrent read-write tasks to evaluate its vertex-grained version control strategy.
% We evaluate different systems on IT and TW in terms of insert throughput and SSSP execution time. 
We evaluate the insertion throughput and execution time of SSSP on different systems in IT and TW.
MBFGraph is not included in the evaluation as it does not support the update-analysis mixed workload.
Initially, we insert 80\% of the edge data as a baseline, followed by continuous edge data insertion while concurrently running the SSSP algorithm.
The results are shown in Fig. \ref{fig:rw}.
\sys outperforms all other systems in the concurrent read-write scenario. 
This is mainly due to the vertex-grained version control of \sys, which is different from the version chain implementation in LSM-trees like RocksDB.
It is more flexible, allowing query requests to access the newly merged data earlier.
Compared to read-only and write-only scenarios, in the concurrent scenario, \sys achieves a 3$\times$ increase in write throughput and a 1.6$\times$ improvement in SSSP performance compared to RocksDB.

%% file: 7-related.tex
\section{Related Work}
\label{sec:relatedwork}

% LSGraph, spruce

\stitle{Dynamic Graph Storage Systems in Memory}. 
There are numerous in-memory dynamic graph storage systems
\cite{sortledton, terrace, aspen, graphone, teseo, pcsr, vcsr, risGraph, livegraph, LSGraph, Spruce}.
Sortleton \cite{sortledton}, GraphOne \cite{graphone}, and others \cite{pcsr, terrace, livegraph, LSGraph, graphone} achieve fast graph data insertion by employing structures with reserved space or linked-list-like structures.
Teseo \cite{teseo}, VCSR \cite{vcsr}, and PCSR \cite{pcsr} adopt a structure similar to CSR and utilize reserved slots to alleviate data migration overhead. 
However, they still suffer from data migration and space expansion when data exceeds the reserved space. 
Spruce \cite{Spruce} is an advanced dynamic in-memory graph structure that allocates a buffer block and a sorted edge block for each vertex to support edge insertions and updates for that vertex.
However, applying this vertex-grained edge block allocation scheme to disk would suffer from poor locality and write amplification
issues, which is similar to LiveGraph \cite{livegraph}. 

\stitle{Dynamic Graph Storage Systems on Disk}.
Several disk-based dynamic graph storage systems have been developed to handle scenarios where the graph data exceeds available memory \cite{livegraph, graphssd, llama, mbfgraph, x-stream, graphchi}. 
LiveGraph \cite{livegraph} supports both in-memory and disk-based scenarios. 
However, LiveGraph's \cite{livegraph} disk performance is poor due to in-place updates and scattered distribution of edge blocks.
GraphSSD \cite{graphssd} and LLAMA \cite{llama} aim to improve disk-based read performance by utilizing structures similar to CSR. 
However, they face challenges in updating their CSRs.
For instance, LLAMA generates multiple snapshots in CSR format to receive updates;
however, read performance degrades when there are too many snapshots.
MBFGraph \cite{mbfgraph} and X-Stream \cite{x-stream} leverage log append approaches to achieve fast write performance and efficient storage of streaming data. 
However, this approach negatively impacts graph analysis performance, as it requires scanning almost all edges of the graph. 
Although MBFGraph uses Bloom filters to reduce the number of edges read, it still suffers from severe I/O amplification.
Unlike these systems, \sys designs a new graph store to combine CSR and LSM-tree to leverage their complementary advantages and simultaneously optimize read and write performance.

\stitle{Graph Databases}.
Numerous graph databases have been developed \cite{NebulaGraph,dgraph2023,neo4j,tugraph,bytegraph,AgensGraph,ArangoDB,JanusGraph,GDB,BG3}.
The main distinction between graph storage systems and graph databases lies in the emphasis of graph databases on transactional support and more complex graph data management. 
Due to the need for transaction support, even simple graph analysis queries like single-source shortest path (SSSP) and PageRank can incur significant overhead in graph databases, resulting in poor performance when running graph analysis algorithms within the database. 
Additionally, these systems support exporting data into various graph data formats (\eg CSR) and running graph analysis algorithms externally. 
However, the cost of the export process alone can exceed the cost of running the algorithms and the freshness of the data is low.
Similar to other graph storage systems, the goal of \sys is to facilitate the fast storage of large-scale graph data and enable graph analysis on the stored data to provide more real-time graph analysis services.

\stitle{Transforming from Static Structure to Dynamic Structure}.
The Bentley–Saxe transformation \cite{statictodyamic1979, statictodyamic1980} is able to transform static data structures into dynamic structures. 
It has been used in various works \cite{dynamicindex,incrementalindex,DynamicSamplingindex}, such as the dynamic extension of static indexes \cite{DynamicSamplingindex}. 
Theoretically, this can be applied to CSR for dynamic updates. 
However, simply transforming CSR into a dynamic structure without considering the disk context, the nature of graph data (\eg power-law distribution), and the characteristics of graph workloads (\eg retrieving all neighbors of a specific vertex) can lead to poor performance in graph updates and analysis.
\sys is designed to provide high read and write performance by leveraging the characteristics of graph data and the common access patterns of graph workloads. Additionally, \sys utilizes LSM-tree to optimize disk I/O overhead caused by random writes.

\stitle{Filter in LSM-tree}.
To support better read performance of LSM-tree,
numerous LSM-trees implementations use filters to test whether a key is contained in a block, thereby reducing unnecessary disk I/O \cite{bloomfilter,Monkey,GRF,Rosetta,CuckooFilter}.
However, the filter of each block needs to be queried, which results in a large number of random memory accesses.
Some novel filters are designed to reduce the number of memory accesses, such as SlimDB \cite{SlimDB}, Chucky \cite{Chucky}, and Mapped SplinterDB \cite{SplinterDB}.
SlimDB \cite{SlimDB} uses a multi-level cuckoo filter \cite{CuckooFilter} to map each key to its level number, and Chucky \cite{Chucky} maintains a filter for the entire LSM-tree, mapping each key to a sub-level number.
Mapped SplinterDB \cite{SplinterDB} uses quotient maplets \cite{qfilter} to map keys to potential SSTables.
However, these methods are not suitable for graph workloads that require to access consecutive vertex IDs, \eg PageRank \cite{page1999pagerank} algorithm. 
For these workloads, they need to test the filters frequently.
GRF \cite{GRF} is a recent global filter that uses full shape encoding for MVCC but relies on a fixed compaction strategy, which is not suitable for \sys, as \sys uses a partial merge leveling strategy to balance read and write performance.

%% file: 8-concl.tex
\section{Conclusion}
\label{sec:concl}
%\vspace{-1ex}
% \noindent
This paper presents {\sys}, a novel dynamic graph storage system that combines the write-friendly LSM-tree and the read-friendly CSR.
We leverage three key designs to enhance system performance and ensure the correctness of read and write tasks.
Firstly, we design an efficient in-memory structure, MemGraph, to enable cache graph updates in memory and persist them to disk efficiently.
Secondly, we design a multi-level CSR equipped with a multi-level index, where the multi-level CSR utilizes the compact CSR structure embedded in the LSM-tree, which can mitigate write amplification and reduce random reads. 
The multi-level index is used to locate the edges of the vertex to reduce the random lookups. 
Lastly, we design a vertex-grained version control mechanism to support concurrent read/write operations and mitigate the impact of compaction on read/write performance.
Our evaluations confirm the efficacy and efficiency of \sys.